\documentclass[a4paper,10pt]{revtex4}
\usepackage{graphicx}  % standard LaTeX graphics tool;
\usepackage{amsmath}   % For getting proper math eqns
\usepackage{amssymb}   % Used for getting various symbols eg. gtrsim, lesssim etc.
\usepackage{bm} % For bold math (esp of lower greek letters)
\usepackage{dcolumn}% Align table columns on decimal point
\usepackage{color}
\usepackage{mathrsfs}
\usepackage{amsfonts}
\usepackage{varioref}
\usepackage{mathrsfs}
\usepackage{graphicx}
\usepackage{latexsym}
\usepackage{amsmath}
\usepackage{amssymb}
\usepackage{textcomp}
\usepackage{amsbsy}
\usepackage{graphics}
\usepackage{epstopdf}
\usepackage{color}
\usepackage[caption=false]{subfig}

\RequirePackage[colorlinks,citecolor=blue,urlcolor=magenta,linkcolor=blue]{hyperref}
\input epsf

\allowdisplaybreaks[4]

\begin{document}

\tolerance=5000

\title{Viable requirements of curvature coupling helical magnetogenesis scenario}

\author{Tanmoy~Paul$^{1}$\,\thanks{pul.tnmy9@gmail.com}} \affiliation{ $^{1)}$ 
Department of Physics, Chandernagore College, Hooghly - 712 136, India }

\begin{abstract}

In the present work, we examine the following points in the context of 
the recently proposed curvature coupling helical magnetogenesis scenario \cite{Bamba:2021wyx} -- (1) whether the model is consistent with the 
predictions of perturbative quantum field theory (QFT), and 
(2) whether the curvature perturbation induced by the generated electromagnetic (EM) field during 
inflation is consistent with the Planck data. Such requirements 
are well motivated in order to argue the viability of the magnetogenesis model under consideration. Actually, the magnetogenesis scenario proposed 
in \cite{Bamba:2021wyx} seems to predict sufficient magnetic strength 
over the large scales and also leads to the correct baryon asymmetry of the universe for a suitable range of the model parameter. 
However in the realm of inflationary magnetogenesis, these 
requirements are not enough to argue the viability of the model, particularly 
one needs to examine some more important requirements in this regard. We may recall that the calculations generally used to determine the magnetic field's 
power spectrum are based on the perturbative QFT -- therefore it is important to examine whether 
the predictions of such perturbative QFT are consistent with the observational bounds of the model parameter. On other hand, the generated gauge field 
acts as a source of the curvature perturbation which needs to be suppressed compared to that of contributed from the inflaton field 
in order to be consistent with the Planck observation. For the perturbative requirement, we examine whether the condition 
$\left|\frac{S_{CB}}{S_{can}}\right| < 1$ is satisfied, where $S_{CB}$ and $S_{can}$ are the non-minimal and the canonical action of the EM field 
respectively. Moreover we determine the power spectrum of the 
curvature perturbation sourced by the EM field during inflation, and evaluate necessary constraints in order to be consistent with the Planck data. 
Interestingly, both the aforementioned requirements in the context of the curvature coupling helical magnetogenesis scenario 
are found to be simultaneously satisfied by that range of the model parameter which leads to the correct magnetic strength 
over the large scale modes.

\end{abstract}

%\pacs{}

\maketitle
\section{Introduction}\label{sec-introduction}

Magnetic fields are observed over a wide range of scales from within galaxy clusters to intergalactic voids 
\cite{Grasso:2000wj,Beck:2000dc,Widrow:2002ud}. From theoretical 
perspective, there are two approaches to understand the origin of such magnetic fields -- (1) the astrophysical origin of the fields which get amplified 
by some dynamo mechanism \cite{Kulsrud:2007an,Brandenburg:2004jv,Subramanian:2009fu} 
and (2) the primordial origin of the magnetic fields from inflationary scenario 
\cite{Jain:2012ga,Durrer:2010mq,Kanno:2009ei,Campanelli:2008kh,
Demozzi:2009fu,Demozzi:2012wh,Bamba:2006ga,Kobayashi:2019uqs,Bamba:2020qdj,Maity:2021qps,Haque:2020bip,
Ratra:1991bn,Ade:2015cva,Chowdhury:2018mhj,Turner:1987bw,
Tripathy:2021sfb,Ferreira:2013sqa,Atmjeet:2014cxa,Kushwaha:2020nfa,Gasperini:1995dh,Giovannini:2021thf,Giovannini:2021xbi,Adshead:2015pva,
Caprini:2014mja,Kobayashi:2014sga,Atmjeet:2013yta,Fujita:2015iga,Campanelli:2015jfa,Tasinato:2014fia} or from the alternative bouncing scenario 
\cite{Frion:2020bxc,Koley:2016jdw,Qian:2016lbf}.

Among all the proposals discussed so far, particularly the inflationary magnetogenesis earned a lot of attention due to its simplicity 
and elegance. Inflation is one of the cosmological scenarios that successfully describes the early stage of the universe, in particular, it 
resolves the flatness and horizon problems, and more importantly, inflation can predict an almost scale invariant 
curvature power spectrum to be well consistent with the recent Planck data \cite{guth,Linde:2005ht,Langlois:2004de,Riotto:2002yw,Baumann:2009ds}. 
So it would be nice if the same inflationary paradigm can 
also describe the origin of the observed magnetic fields, which is the essence of inflationary magnetogenesis. However in the 
standard Maxwell's theory, the electromagnetic (EM) field does not fluctuate over the vacuum state due to the conformal invariance of the EM action, 
and thus a sufficient amount of magnetic field can not be generated at present epoch of the universe. The way to boost the magnetic energy 
from the vacuum state is to break the conformal invariance of the EM action, and this can be suitably done by introducing 
a non-minimal coupling of the EM field with the background inflaton field or with the background spacetime curvature 
\cite{Jain:2012ga,Durrer:2010mq,Kanno:2009ei,Campanelli:2008kh,
Demozzi:2009fu,Demozzi:2012wh,Bamba:2006ga,Kobayashi:2019uqs,Bamba:2020qdj,Maity:2021qps,Haque:2020bip,
Ratra:1991bn,Ade:2015cva,Chowdhury:2018mhj,Turner:1987bw,
Tripathy:2021sfb,Ferreira:2013sqa,Atmjeet:2014cxa,Kushwaha:2020nfa,Giovannini:2021thf,Adshead:2015pva,
Caprini:2014mja,Kobayashi:2014sga,Atmjeet:2013yta,Fujita:2015iga,Campanelli:2015jfa,Tasinato:2014fia}. 
Moreover depending on the nature of the electromagnetic coupling function, the parity symmetry of the EM field 
may or may not be violated and thus the EM field can have either helical or non-helical respectively. However this simple way of inflationary magnetogenesis may be riddled with some problems, like the backreaction issue 
and the strong coupling problem. The backreaction issue arises when the EM field energy density dominates (or becomes comparable) over the 
background energy density, which in turn spoils the background inflationary expansion of the universe. On other hand, the strong coupling problem is related 
when the effective electric charge becomes strong during inflation. Therefore the backreaction and the strong coupling problems 
need to be resolved in a successful inflationary magnetogenesis scenario (see \cite{Demozzi:2009fu,Ferreira:2013sqa,Tasinato:2014fia,Nandi:2021lpf}). 
Besides during the inflation, 
the occurrence of a prolonged reheating phase after the inflation has been proved to play a significant role in 
magnetic field's power spectrum (for studies of various reheating mechanisms, see 
\cite{Dai:2014jja,Cook:2015vqa,Albrecht:1982mp,Ellis:2015pla,Ueno:2016dim,Eshaghi:2016kne,Maity:2018qhi,Haque:2021dha,DiMarco:2017zek,Drewes:2017fmn}). 
Such effects of the reheating phase having non-zero e-fold number in the realm of inflationary magnetogenesis 
have been addressed in the context of curvature coupling as well as scalar coupling magnetogenesis scenario 
\cite{Bamba:2021wyx,Kobayashi:2019uqs,Bamba:2020qdj,Maity:2021qps,Haque:2020bip}. 
Actually the existence of a strong electric field at the end of 
inflation induces the magnetic field during the reheating phase from Faraday's law of induction, which in turn enhances the magnetic strength at 
current epoch. 

Recently we have proposed a curvature coupling helical magnetogenesis model where the conformal and parity symmetries 
of the electromagnetic field are broken through its non-minimal coupling to the background $f(R,\mathcal{G})$ gravity 
via the dual field tensor, so that the generated magnetic field is helical in nature \cite{Bamba:2021wyx}. This is well motivated from the rich cosmological 
consequences of $f(R,\mathcal{G})$ gravity, see \cite{Nojiri:2005vv,Li:2007jm,Carter:2005fu,Nojiri:2019dwl,
Cognola:2006eg,Chakraborty:2018scm,Elizalde:2020zcb,Nojiri:2022xdo,Odintsov:2022unp} for various perspectives of $f(R,\mathcal{G})$ cosmology. 
After the end of inflation, the universe 
enters to a reheating phase and depending on the reheating mechanism, we have considered two different reheating scenarios 
in \cite{Bamba:2021wyx}, namely -- (a) the instantaneous 
reheating where the universe instantaneously converts to the radiation era immediately after the inflation, 
and (b) the Kamionkowski reheating scenario characterized by a non-zero reheating e-fold number and a constant equation of state parameter. The proposed 
magnetogenesis scenario shows the following features: (1) for both the reheating cases, the model predicts sufficient magnetic strength over the large scale 
modes at present universe for a suitable range of the model parameter; (2) 
the model is free from the backreaction and the strong coupling problems; (3) due to the helical nature, the magnetic field of strength 
$10^{-13}\mathrm{G}$ over the galactic scales predicts the correct baryon asymmetry of the universe that is consistent with the observation. However 
in the realm of inflationary magnetogenesis, these requirements are not enough to argue the viability of a magnetogenesis model, in particular, 
one needs to examine some more important requirements in order to argue the viability of the model. In this regard, one may recall that 
the calculations that we use to determine the magnetic field's evolution and its power spectrum are based on the perturbative quantum field theory -- 
therefore it is important to examine whether the predictions of such perturbative QFT are consistent with the observational bounds of the model parameter. 
Such perturbative requirement in the context of axion magnetogenesis scenario was studied earlier in \cite{Durrer:2010mq,Ferreira:2015omg}. On 
other hand, the generated EM field may source the curvature perturbation during inflation at super-Hubble scales. 
Therefore, by considering that the curvature perturbation observed through the Planck data is mainly contributed from the slow-roll 
inflaton field, we need to investigate whether the curvature perturbation induced by the EM field does not exceed than that of induced 
by the background inflaton field in order to be consistent with the recent Planck observation. The authors of 
\cite{Fujita:2013qxa,Barnaby:2012tk,Bamba:2014vda,Suyama:2012wh} addressed the induced curvature 
perturbation from the EM field and determined the necessary constraints in scalar coupling inflationary magnetogenesis scenario. However in the context 
of curvature coupling magnetogenesis scenario, the investigation of 
such perturbative requirement and the induced curvature perturbation from the EM field have not 
yet given proper attention.

Motivated by the above arguments, in the present work, we will study the following points in 
the curvature coupling helical magnetogenesis model proposed in \cite{Bamba:2021wyx} :

\begin{itemize}
 \item Is the model consistent with the perturbative requirement ?
 
 \item What about the power spectrum for the curvature perturbation sourced by the EM field during inflation ? 
 Is it compatible with the Planck observation ?
\end{itemize}

For the perturbative requirement, we will examine whether the condition $\left|\frac{S_{CB}}{S_{can}}\right| < 1$ is satisfied, where $S_{can}$ 
and $S_{CB}$ are the canonical and the conformal breaking action of the EM field respectively. This condition indicates that the loop contribution 
in the EM two-point correlator is less than that the tree propagator of the EM field, as the loop 
contribution in the EM propagator arises due to the presence of the action $S_{CB}$. In regard to the second requirement, we will calculate the 
power spectrum of the curvature perturbation induced by the EM field during inflation 
and will determine the necessary constraints in order to have a consistent model with the Planck data. The model parameter(s) 
will be critically scanned so that both the above requirements, along with the large scale observations of magnetic field, are concomitantly satisfied.

The paper is organized as follows: in Sec.[\ref{sec-model}], we will briefly describe the essential features of the 
magnetogenesis model that we will use in the present work. In Sec.[\ref{sec-cut-off}], Sec.[\ref{sec-perturbative}] and 
Sec.[\ref{sec-perturbation}], we will determine the cut-off scale, the perturbative requirement and the induced curvature perturbation of the model 
respectively, and will reveal the necessary constraints. The paper ends with some conclusions. Finally we would like to clarify the notations and conventions 
that we will use in the subsequent calculations. We will work with an isotropic and homogeneous Freidmann Robertson Walker (FRW) spacetime where the 
metric is:
\begin{eqnarray}
 ds^2 = -dt^2 + a^2(t)\delta_{ij}dx^{i}dx^{j}\nonumber
\end{eqnarray}
with $a(t)$ being the scale factor of the universe and $t$ is the cosmic time. The conformal time and the e-folding number 
will be denoted by $\eta$ and $N$ respectively. An overdot 
and an overprime will indicate $\frac{d}{dt}$ and $\frac{d}{d\eta}$ respectively. A quantity with a suffix 'f' will represent the quantity at the end of 
inflation, for example, $N_\mathrm{f}$ is the total inflationary e-folding number, $k_f$ represents the mode that crosses 
the Hubble horizon at the end of inflation etc. Moreover 
the cosmic Hubble parameter will be symbolized by $H = \dot{a}/a$ and the conformal Hubble parameter 
will be $\mathcal{H} = a'/a$.

\section{Essential features of the magnetogenesis model}\label{sec-model}

Here we consider the higher curvature helical magnetogenesis scenario that we proposed in 
\cite{Bamba:2021wyx} where the electromagnetic dual field tensor couples with the background Ricci scalar as well as with the Gauss-Bonnet scalar. 
The action is given by,
\begin{eqnarray}
 S = S_{grav} + S_{em}^{(can)} + S_{CB}~~.
 \label{action0}
\end{eqnarray}
where $S_{grav}$ is the gravitational action that serves the inflationary agent during the early universe, and is given by
\begin{eqnarray}
 S_{grav} = \int d^4x \sqrt{-g}~\mathcal{F}(\Phi,R,\mathcal{G})~~.\label{1}
\end{eqnarray}
Here $\Phi$ is a scalar field under consideration, $R$ and $\mathcal{G}$ are the background Ricci scalar and the background Gauss-Bonnet terms 
respectively. At this stage, we do not propose any particular form of $\mathcal{F}(\Phi,R,\mathcal{G})$ for the background gravitational action. 
Actually we will give some suitable forms of $\mathcal{F}(\Phi,R,\mathcal{G})$ which lead to successful inflation, and thus, 
any of such forms of $\mathcal{F}(\Phi,R,\mathcal{G})$ is allowed in the context of magnetogenesis scenario. 
In this work we consider power law inflationary scenario to evaluate the power spectrum of the electromagnetic fluctuations. For 
power law inflation, the scale factor is given by $a(t) \propto t^{p}$ with $p > 1$. In the conformal time (symbolized by 
$\eta$), the scale factor reads as \cite{Shankaranarayanan:2004iq}
\begin{eqnarray}
 a(\eta) = \bigg(\frac{-\eta}{\eta_0}\bigg)^{\beta + 1}~~~~~~~~\mathrm{where}~~~~~~~~~\beta = -\left(\frac{2p-1}{p-1}\right)~~,
 \label{scale factor}
\end{eqnarray}
and $\eta_0$ is a constant having mass dimension $=[-1]$, and $\eta_0^{-1}$ denotes the scale of inflation. 
Moreover an overprime denotes $\frac{d}{d\eta}$ and $\mathcal{H}$ is the conformal Hubble parameter defined by $\mathcal{H} = a'/a$. Using the 
above expression of $a(\eta)$, we get,
\begin{eqnarray}
 \mathcal{H} = \frac{\beta + 1}{\eta}~~.
 \label{Hubble parameter}
\end{eqnarray}
In the subsequent calculations, the e-folding number will be represented by $N$, and $N = 0$ indicates the beginning of inflation, i.e the e-folding 
number is increasing as the inflation goes on. For the above scale factor, the cosmic Hubble parameter 
(defined by $H = \frac{\dot{a}}{a}$ with an overdot symbolizes the derivative with respect to cosmic time $t$) is given by,
\begin{eqnarray}
 H = H_0\exp{\left(-\frac{\delta}{1+\delta} N\right)}~~~~~~~~\mathrm{with}~~~~~~~~\delta = -\beta - 2 = \frac{1}{p-1},
 \label{cosmic Hubble parameter}
\end{eqnarray}
in terms of the e-folding number, where $H_0$ is a constant that represents the Hubble parameter at the beginning of inflation. 
Here we would like to mention that for the scale factor of Eq.(\ref{scale factor}), the slow roll parameter comes as $\epsilon = 1/p$ and thus 
$\epsilon \neq \delta$. However due to $p > 1$, the slow roll parameter is slightly different than $\delta$, for example, $p = 11$ leads to 
$\epsilon \simeq 0.09$ and $\delta = 0.1$. 

Now we will propose some suitable forms of $\mathcal{F}(\Phi,R,\mathcal{G})$ which indeed leads to power law inflation:
\begin{itemize}
 \item The action with a non-minimally coupled scalar field, where the $\mathcal{F}(\Phi,R,\mathcal{G})$ is given by \cite{delCampo:2015wma},
 \begin{eqnarray}
  \mathcal{F}(\Phi,R,\mathcal{G}) = \left(\frac{1}{16\pi G} + \xi\Phi^2\right)R - \frac{1}{2}g^{\mu\nu}\partial_{\mu}\Phi\partial_{\nu}\Phi - V(\Phi)~~,
  \label{1m}
 \end{eqnarray}
results to a viable power law inflation described by $a(t) \propto t^{p}$ with $p>1$. Here $G$ is the Newton's constant, 
$\xi$ is the non-minimal coupling of the scalar field and $V(\Phi)$ is the scalar field 
potential which has the following form,
\begin{eqnarray}
 V(\Phi) = \left(\Phi + \gamma\right)^{1-2/n}\left(A\Phi - \gamma B\right)~~,\nonumber
\end{eqnarray}
where $\gamma$, $A$, $B$ are constants and $n$ is related to the exponent of the scale factor ($p$) as $p = \frac{n^2 - n}{2+n}$. The authors 
of \cite{delCampo:2015wma} showed that the inflationary quantities lie within the observational constraints for $10.04 \lesssim p \lesssim 15.03$.

\item The f(R) model given by \cite{Sharma:2022tce},
\begin{eqnarray}
 \mathcal{F}(\Phi,R,\mathcal{G}) \propto R^{1+\sigma}~~,
 \label{2}
\end{eqnarray}
allows a power law inflationary solution $a(t) \propto t^{p}$ (with $p>1$) when $p$ and $\sigma$ are related by 
$p = \frac{\sigma(1+2\sigma)}{(1-\sigma)}$. 
It has been shown in \cite{Sharma:2022tce} that the inflationary quantities in the context of such power law inflation 
satisfy the recent Planck constraints for $10.85 \leq p \leq 12.45$.

\item In the context of k-Gauss-Bonnet inflation, the Gauss-Bonnet term gets coupled with the kinetic term of a scalar field under consideration. In 
particular, the $\mathcal{F}(\Phi,R,\mathcal{G})$ is given by \cite{Pham:2021fjj},
\begin{eqnarray}
 \mathcal{F}(\Phi,R,\mathcal{G}) = \frac{R}{16\pi G} + X - \frac{1}{8}J(X)\mathcal{G}~~,
 \label{3}
\end{eqnarray}
where $X$ is the kinetic term of the scalar field. A stable power law inflationary solution of the form $a(t) \propto t^{p}$ (with $p>1$) 
can be obtained from the above model for $J(X) \propto X^{n}$, where $n$ and $p$ are related by a suitable fashion given in \cite{Pham:2021fjj}. Here 
it deserves mentioning that in absence of scalar field potential, the power law inflation in the k-Gauss-Bonnet model leads to the stability 
of the primordial tensor perturbation \cite{Pham:2021fjj}.
\end{itemize}

Based on the above arguments, if we consider the the background action of Eq.(\ref{1m}) then the exponent 
of the power law inflationary scale factor should lie within $10.04 \lesssim p \lesssim 15.03$, or, if we consider the gravitational action of 
Eq.(\ref{2}) then we need to choose $10.85 \leq p \leq 12.45$ -- in order to get a viable power law inflation. Keeping this in mind, 
we consider $p = 11$ in the present context, for which, one gets $\beta = -2.1$ or $\delta = 0.1$ or $\epsilon \simeq 0.09$ 
(see Eq.(\ref{scale factor}) and Eq.(\ref{cosmic Hubble parameter}) for the expressions of $\beta$ and $\delta$ respectively). 
We will demonstrate that with this value of $\delta$, the 
current magnetogenesis scenario predicts sufficient magnetic strength for suitable values of other model parameters.\\

The $S_{em}^{(can)}$ and $S_{CB}$ in Eq.(\ref{action0}) are the canonical kinetic term and the non-minimal coupling 
of the EM field respectively. In particular, 
\begin{eqnarray}
 S_{em}^{(can)} = \int d^4x\sqrt{-g}\big[-\frac{1}{4}F_{\mu\nu}F^{\mu\nu}\big]~~,
 \label{action part 2}
\end{eqnarray}
and
\begin{eqnarray}
 S_{CB} = \int d^4x\sqrt{-g} f(R,\mathcal{G})\left[-\lambda F_{\mu\nu}\widetilde{F}^{\mu\nu}\right]~~,
 \label{action part 3}
\end{eqnarray}
respectively. Here $F_{\mu\nu} = \partial_{\mu}A_{\nu} - \partial_{\nu}A_{\mu}$ represents the EM field tensor and $A_{\mu}$ is the corresponding 
EM field. Moreover $\widetilde{F}^{\mu\nu} = \epsilon^{\mu\nu\alpha\beta}F_{\alpha\beta}$ where 
$\epsilon^{\mu\nu\alpha\beta}$ is the four dimensional Levi-Civita tensor defined by $\epsilon^{\mu\nu\alpha\beta} 
= -\frac{1}{\sqrt{-g}}[\mu\nu\alpha\beta]$, the $[\mu\nu\alpha\beta]$ symbolizes the completely antisymmetric permutation with $[0123] = 1$. 
Eq.(\ref{action part 3}) reveals that the EM field couples with the background Ricci scalar as well as with the Gauss-Bonnet scalar through the 
non-minimal coupling function $f(R,\mathcal{G})$. The form of $f(R,\mathcal{G})$ is considered to be a power law of $R$ and $\mathcal{G}$, particularly
\begin{eqnarray}
 f(R,\mathcal{G}) = \kappa^{2q}\big(R^q + \mathcal{G}^{q/2}\big)~~,
 \label{form of f 1}
\end{eqnarray}
with $q$ being a parameter of the model and $\kappa = M_\mathrm{Pl}^{-1} = \sqrt{8\pi G}$, where $G$ is the Newton's constant. 
The parameter $q$ plays an important role in regard to the estimation of magnetic field 
at current universe. The presence of $S_{CB}$ spoils the conformal invariance, however preserves the U(1) symmetry, 
of the EM action. Furthermore, 
Eq.(\ref{action part 3}) depicts that the EM field couples with the background spacetime curvature via its dual tensor 
($F\widetilde{F}$), which further breaks the parity symmetry of the EM field, and consequently, the generated EM field turns out to be 
helical in nature. With Eq.(\ref{scale factor}) and Eq.(\ref{Hubble parameter}), the explicit form of $f(R,\mathcal{G})$ from 
Eq.(\ref{form of f 1}) becomes,
\begin{eqnarray}
 f(R,\mathcal{G}) = \kappa^{2q}\bigg\{\frac{\big[6\beta(\beta+1)\big]^q 
 + \big[-24(\beta+1)^3\big]^{q/2}}{\eta_0^{2q}}\bigg\}\bigg(\frac{-\eta}{\eta_0}\bigg)^{2\delta q}~~.
 \label{form of f 2}
\end{eqnarray}
Varying the action Eq.(\ref{action0}) with respect to $A_{\mu}$, we get
\begin{eqnarray}
 \partial_{\alpha}\left[\sqrt{-g}\left\{g^{\mu\alpha}g^{\nu\beta}F_{\mu\nu} 
 + 8\lambda f(R,\mathcal{G})~\epsilon^{\mu\nu\alpha\beta}\partial_{\mu}A_{\nu}\right\}\right] = 0~~.
 \label{eom}
\end{eqnarray}
We will work with the Coulomb gauge i.e $A_0 = 0$ and $\partial_{i}A^{i} = 0$, due to which, the temporal component of 
Eq.(\ref{eom}) becomes trivial, while the spatial component of the same becomes,
\begin{eqnarray}
 A_i''(\eta,\vec{x}) - \partial_l\partial^{l}A_i + 8\lambda f'(R,\mathcal{G})~\epsilon_{ijk}\partial_{j}A_{k} = 0~~,
 \label{eom2}
\end{eqnarray}
where $\epsilon_{ijk} = \left[0ijk\right]$ and $f'(R,\mathcal{G}) = \frac{df}{d\eta}$. It is evident that the presence of the $f(R,\mathcal{G})$ 
modifies the EM field equation in comparison to the standard Maxwell's equation. At this stage we quantize the EM field, so that one does not need 
an initial seed magnetic field at classical level, and we may argue that the EM field generates from the quantum vacuum state. For this purpose, we use,
\begin{eqnarray}
 \hat{A}_i(\eta,\vec{x}) = \int \frac{d\vec{k}}{(2\pi)^3}\sum_{r=+,-}\epsilon_{ri}~\bigg[\hat{b}_r(\vec{k})A_{r}(k,\eta)e^{i\vec{k}.\vec{x}} 
 + \hat{b}_r^{+}(\vec{k})A_{r}^{*}(k,\eta)e^{-i\vec{k}.\vec{x}}\bigg]~~,
 \label{mode decomposition}
\end{eqnarray}
where $\vec{k}$ is the EM wave vector, $r = +,-$ runs along the polarization index with $\vec{\epsilon}_{+}$ and 
$\vec{\epsilon}_{-}$ are two polarization vectors and $A_r(k,\eta)$ is the $k$-th mode function for the EM field. In the present context, since 
the magnetic field is helical in nature, we work with the helicity basis set where the 
polarization vectors are given by: $\vec{\epsilon}_{+} = \frac{1}{\sqrt{2}}\left(1,i,0\right)$ and 
$\vec{\epsilon}_{-} = \frac{1}{\sqrt{2}}\left(1,-i,0\right)$ respectively. Consequently, $A_{\pm}(k,\eta)$ follows:
\begin{eqnarray}
 A_{\pm}''(k,\eta) + \left[k^2 \mp k\left(\frac{\zeta^2}{\eta^2}\right)\left(\frac{-\eta_0}{\eta}\right)^{2\alpha}\right]A_{\pm}(k,\eta) = 0~~,
 \label{FT eom3}
\end{eqnarray}
where $\zeta^2$ and $\alpha$ have the following forms,
\begin{eqnarray}
 \zeta^2&=&\left(16\delta q\lambda\eta_0\right)\left(\frac{\kappa}{\eta_0}\right)^{2q}\bigg\{\big[6\beta(\beta+1)\big]^q 
 + \big[-24(\beta+1)^3\big]^{q/2}\bigg\}~,\nonumber\\
 \alpha&=&-\frac{1}{2} - \delta q~~.
 \label{B}
\end{eqnarray}
Therefore the photon dispersion relation in the present context is given by,
\begin{eqnarray}
 \omega_\mathrm{\pm}^2 = k^2 \mp k.~\mathrm{constant}/\eta^{1-2\delta q}~~,\nonumber
\end{eqnarray}
which, due to the presence of the factor '$\delta q$', 
is different than the axion magnetogenesis like model where a (pseudo) scalar field gets coupled linearly with the Chern-Simons term 
\cite{Anber:2006xt,Barnaby:2011vw,Peloso:2016gqs}. We will show below that the presence of $\delta q$ is crucial, 
due to which, the present curvature coupled magnetogenesis scenario 
predicts sufficient magnetic strength at the current universe.

In the sub-Hubble scale when the relevant modes lie within 
the Hubble horizon, one can neglect the term containing $\zeta^2$ in Eq.(\ref{FT eom3}), 
and thus both the EM mode functions remain in the Bunch-Davies vacuum state. However 
in the super-Hubble scale when the modes get outside from the Hubble horizon, the term containing $\zeta^2$ in Eq.(\ref{FT eom3}) dominates over the 
$k^2$ term, and thus $A_{\pm}(k,\eta)$ has the following solution in the super-Hubble scale,
\begin{eqnarray}
 A_+(k,\eta)&=&\left(\frac{C_1 - C_2\cot{\left(\frac{-\pi}{2\alpha}\right)}}{\Gamma\left(1 + \frac{1}{2\alpha}\right)}\right)
 \left(-i\frac{\zeta\sqrt{k}}{2\alpha}\right)^{1/(2\alpha)}~~,\nonumber\\
 A_-(k,\eta)&=&\left(\frac{C_3 - C_4\cot{\left(\frac{-\pi}{2\alpha}\right)}}{\Gamma\left(1 + \frac{1}{2\alpha}\right)}\right)
 \left(\frac{\zeta\sqrt{k}}{2\alpha}\right)^{1/(2\alpha)}~~.
 \label{superhorizon form 1}
 \end{eqnarray}
Here $C_i$ ($i=1,2,3,4$) are integration constants that can be determined from the Bunch-Davies initial condition, the explicit forms of 
$C_i$ are shown in the Appendix (Sec.[\ref{sec-app}]). In the expressions of $C_1$ and $C_2$ the arguments inside the Bessel functions are 
complex, unlike to that of $C_3$ and $C_4$ where the Bessel functions contain real arguments. This makes $C_1 \approx C_2 \gg C_3 \approx C_4$, or 
equivalently $A_{+}(k,\eta) \gg A_{-}(k,\eta)$, i.e the amplitude of the positive helicity mode during inflation is much larger 
than that of the negative helicity mode. Consequently $A_{\pm}'(k,\eta)$ are given by,
\begin{eqnarray}
 \frac{dA_+}{d(-k\eta)}&=&\left(\frac{H_0}{k}\right)\left(\frac{C_2\Gamma\left(\frac{1}{2\alpha}\right)}{\pi}\right)
 \left(-i\frac{\zeta\sqrt{k}}{2\alpha}\right)^{-1/(2\alpha)}~~,\nonumber\\
 \frac{dA_-}{d(-k\eta)}&=&\left(\frac{H_0}{k}\right)\left(\frac{C_4\Gamma\left(\frac{1}{2\alpha}\right)}{\pi}\right)
 \left(\frac{\zeta\sqrt{k}}{2\alpha}\right)^{-1/(2\alpha)}~~.
 \label{superhorizon form 2}
\end{eqnarray}
With the above expressions of $A_{\pm}'(k,\eta)$ and $A_{\pm}(k,\eta)$, the electric and magnetic power spectra 
during inflation are given by \cite{Bamba:2021wyx},
\begin{eqnarray}
 \mathcal{P}(\vec{E}) = \frac{k}{2\pi^2}~\frac{k^2}{a^4}\big|A_{+}'(k,\eta)\big|^2 = 
 \left(\frac{k}{2\pi^4}\right)\left(\frac{H_0}{k}\right)^2\left(\frac{k}{a}\right)^4
 \left|\left(\frac{\zeta\sqrt{k}}{2\alpha}\right)^{-\frac{1}{2\alpha}}\Gamma\left(\frac{1}{2\alpha}\right)\right|^2
 \left\{\left|C_2\right|^2\right\}
 \label{electric power spectrum 1}
\end{eqnarray}
and
 \begin{eqnarray}
 \mathcal{P}(\vec{B}) = \frac{k}{2\pi^2}~\frac{k^4}{a^4}\big|A_+(k,\eta)\big|^2 = 
 \left(\frac{k}{2\pi^2}\right)\left(\frac{k}{a}\right)^4\left|\frac{\left(\frac{\zeta\sqrt{k}}{2\alpha}\right)^{\frac{1}{2\alpha}}}
 {\Gamma\left(1 + \frac{1}{2\alpha}\right)}\right|^2
 \left\{\left|C_1 - C_2\cot{\left(-\frac{\pi}{2\alpha}\right)}\right|^2\right\}
 \label{magnetic power spectrum 1}
\end{eqnarray}
respectively, where we consider the contribution from the positive helicity mode only, due to $A_{+}(k,\eta) \gg A_{-}(k,\eta)$. 
It is evident that both the $\mathcal{P}(\vec{E})$ and $\mathcal{P}(\vec{B})$ tend to zero as $|k\eta| \rightarrow 0$ (i.e 
near the end of inflation), which indicates that the EM field has negligible backreaction on the background spacetime 
(for detailed analysis of the backreaction issue in the present magnetogenesis model, see \cite{Bamba:2021wyx}). Moreover the helicity power spectrum 
during the inflation is given by,
\begin{eqnarray}
 \mathcal{P}_h 
 = \frac{k}{2\pi^2}~\frac{k^3}{a^3}\big|A_{+}(k,\eta)\big|^2 = 
 \left(\frac{k}{2\pi^2}\right)\left(\frac{k}{a}\right)^3\left|\frac{\left(\frac{\zeta\sqrt{k}}{2\alpha}\right)^{1/(2\alpha)}}
 {\Gamma\left(1 + \frac{1}{2\alpha}\right)}\right|^2
 \left\{\left|C_1 - C_2\cot{\left(-\frac{\pi}{2\alpha}\right)}\right|^2\right\}~~.
 \label{helicity power spectrum inflation}
\end{eqnarray}

After the inflation ends, the universe enters to a reheating phase and depending on the reheating mechanisms, we consider two different reheating 
scenarios -- (a) instantaneous reheating, in which case, the universe instantaneously converts to the radiation era immediately after the inflation, and 
hence the e-folding number of the instantaneous reheating is zero; 
(b) the Kamionkowski reheating proposed in \cite{Dai:2014jja}, which has a non-zero e-fold number and characterized 
by a reheating equation of state (EoS) parameter ($\omega_\mathrm{eff}$) and a reheating temperature ($T_\mathrm{re}$). In the instantaneous reheating case, 
the magnetic field energy density redshifts by $a^{-4}$ from the end of inflation to the present epoch. However in the Kamionkowski reheating case, 
the scenario becomes different, in particular, the magnetic energy density follows a non-trivial evolution during the reheating phase and then 
goes by the usual redshift $a^{-4}$ from the end of reheating to the present epoch of the universe. During the Kamionkowski reheating era, 
the magnetic power spectrum is controlled by the two factors: $a^{-4}$ and $(a^3H)^{-2}$ respectively 
($H$ is the Hubble parameter during the reheating era), where the later factor encodes the information of the prolonged reheating stage. At this stage 
it deserves mentioning that the effect of $(a^3H)^{-2}$ depends on the hierarchy between the electric and the magnetic field at the end of inflation. 
In particular, if the electric field at the end of inflation becomes much stronger that that of the magnetic field (nearly 
$\frac{\mathcal{P}(\vec{E})}{\mathcal{P}(\vec{B})} \sim e^{2N_\mathrm{f}}$ where $N_\mathrm{f}$ is the total inflationary e-fold number), the 
effect of $(a^3H)^{-2}$ becomes dominant over the other one, and then the reheating phase shows an important role in the magnetic field's evolution.

In the present context of higher curvature helical magnetogenesis scenario, we showed that -- (1) the EM field has negligible backreaction on the 
background spacetime and does not jeopardize the inflationary expansion, (2) the model is free from the strong coupling problem, 
(3) for both the reheating cases, the model predicts sufficient 
magnetic strength at current epoch of the universe for a suitable range of $q$ given by: $2.1 \leq q \leq 2.26$ for the instantaneous reheating scenario and 
$2.1 \leq q \leq 2.25$ for the Kamionkowski reheating case respectively \cite{Bamba:2021wyx}, and 
(4) due to the helical nature, the magnetic field of strength 
$10^{-13}\mathrm{G}$ over the galactic scales predicts the correct baryon asymmetry of the universe that is consistent with the observation. 
Here we would like to mention that related results of baryogenesis can be obtained 
when the EM field dual tensor couples to an axion field with cosmological time dependence, 
that leads to tachyonic instabilities and results to a growth of magnetic field \cite{Guendelman:1991se}. 
It is evident that the viable range of $q$ is almost same for both the 
reheating cases. This is due to the reason that the electric and the 
magnetic field do not have enough hierarchy at the end of inflation, 
which in turn makes the instantaneous and Kamionkowski reheating scenarios almost similar in respect to the 
EM field's evolution.\\

Thus as a whole, the present magnetogenesis model with $q = [2.1,2.25]$ is found to be viable 
in regard to the CMB observations of the current magnetic field as well as free from the backreaction and the strong coupling issues. 
However these requirements are 
not sufficient to argue that a magnetogenesis model is a 
viable model, particularly we need to investigate some more important requirements in this regard. 
Here one needs to recall that the calculations regarding the magnetic field's evolution and its power spectrum 
are based on perturbative QFT -- therefore it is important to examine whether the magnetogenesis model under consideration 
is consistent with the predictions of such perturbative QFT. On other hand, the generation of primordial EM field may source 
the curvature perturbation in the super-Hubble scales, and thus 
we need to investigate whether the curvature perturbation induced by the EM field does not exceed than the curvature perturbation contributed from 
the background inflaton field in order to be consistent with the Planck data. 
Thus in the present higher curvature helical magnetogenesis scenario, our aim is to investigate the following 
points -- (a) whether the underlying theory of the model is consistent with perturbative QFT, 
and (b) whether the curvature perturbation induced by the EM field 
does not exceed than that of coming from the inflaton field. 
As mentioned earlier that the range $q=[2.1,2.25]$ leads to the correct magnetic field 
in the present context, thus we will examine the above mentioned requirements in this range of $q$ in order to keep intact the generation of EM field.

However before moving to examine the perturbative validity, we first determine the cut-off scale of the present model 
by using the power counting analysis as demonstrated in \cite{Burgess:2009ea,Hertzberg:2010dc,Bezrukov:2010jz}, and check whether the relevant 
energy scales lie below the cut-off scale. This in turn will provide a hint for the perturbative validity of the model.

\section{The cut-off scale of the model}\label{sec-cut-off}

\begin{figure}[!h]
\begin{center}
 \centering
 \includegraphics[width=3.4in,height=2.5in]{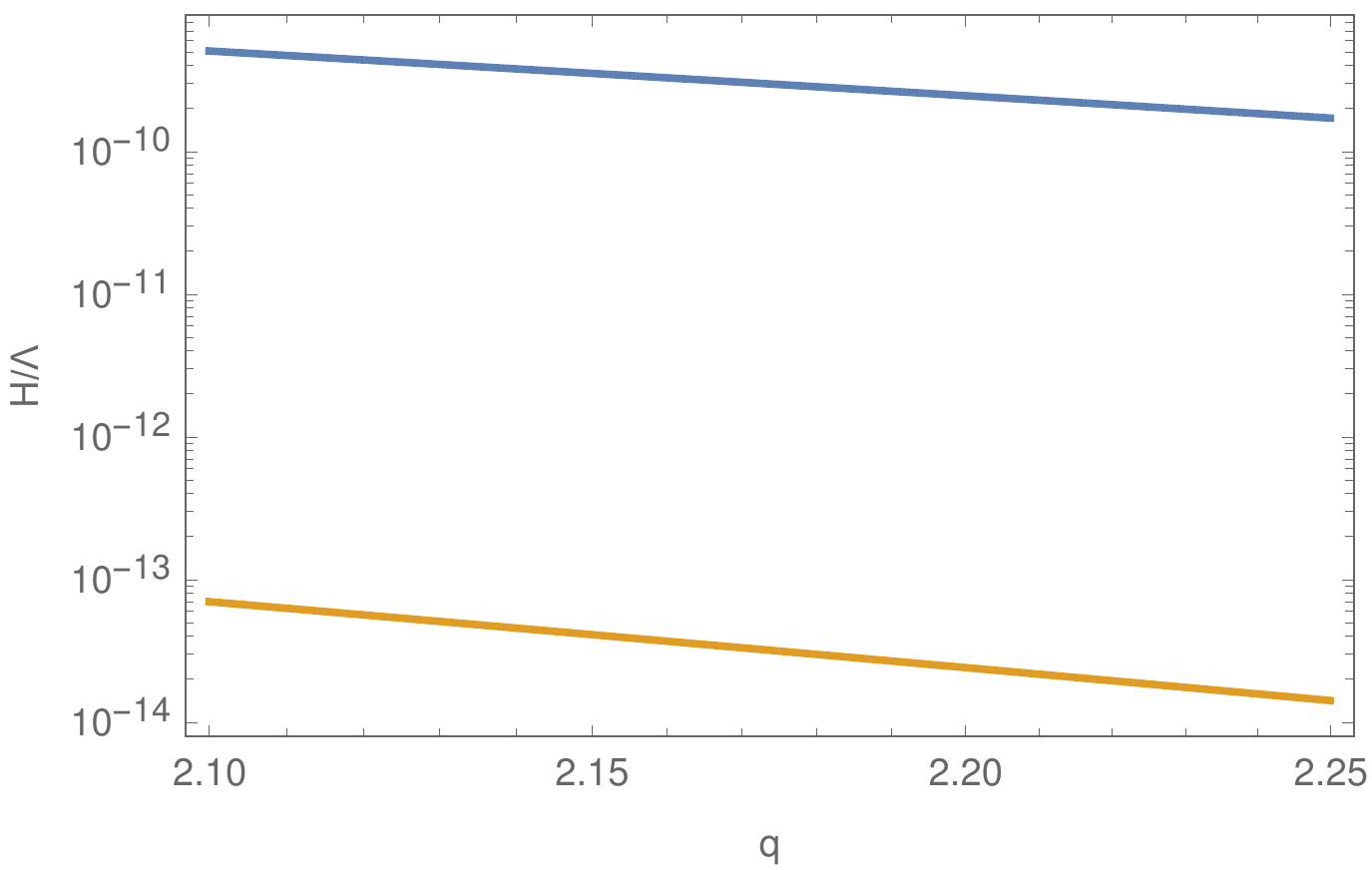}
 \caption{$H/\Lambda$ versus $q$ in the range $2.1 \leq q \leq 2.25$, with $\lambda = 1$, $\delta = 0.1$, $H_0 = 10^{13}\mathrm{GeV}$ 
 and $N_\mathrm{f}=51$. The blue curve represents the ratio of $\frac{H}{\Lambda}$ 
 at the beginning of inflation when $H = H_0$, and the yellow curve specifies $\frac{H}{\Lambda}$ at the end of inflation 
 when $H =  H_0\exp{\left[-\left(\frac{\delta}{\delta+1}\right) N_\mathrm{f}\right]}$.} 
 \label{plot-eft}
\end{center}
\end{figure}

To estimate the cut-off scale, we expand the metric around the background FRW spacetime,
\begin{eqnarray}
 g_{\mu\nu} = \overline{g}_{\mu\nu} + \kappa h_{\mu\nu}~~,
 \label{eft 1}
\end{eqnarray}
where $\overline{g}_{\mu\nu}$ is the FRW metric and 
$h_{\mu\nu}$ are metric perturbations with mass dimension $=[+1]$. Consequently, the determinant of the metric gets 
the following expressions ( in the leading order of $\mathcal{O}\big(\kappa h_{\mu\nu}\big)$ ) around its background value,
\begin{eqnarray}
 \sqrt{-g} = \sqrt{-\overline{g}}\left\{1 + \frac{\kappa}{2}h^{\mu}_{\mu}\right\}~~.
 \label{eft 2}
\end{eqnarray}
The variation of Ricci scalar and the Gauss-Bonnet scalar are given by,
\begin{eqnarray}
 \delta R = \kappa\left\{-h_{\mu\nu}\overline{R}^{\mu\nu} + \overline{\nabla}^{\mu}\overline{\nabla}^{\nu}h_{\mu\nu} - \overline{\Box}h^{\mu}_{\mu}\right\}~~,
 \label{eft 3}
\end{eqnarray}
and
\begin{eqnarray}
 \delta \mathcal{G}&=&2\kappa \overline{R}\left\{-h_{\mu\nu}\overline{R}^{\mu\nu} + \overline{\nabla}^{\mu}\overline{\nabla}^{\nu}h_{\mu\nu} 
 - \overline{\Box}h^{\mu}_{\mu}\right\} 
 + 2\kappa\left\{4\overline{R}^{\rho\sigma}\overline{R}^{\mu~\nu}_{~\rho~\sigma} - \overline{R}^{\mu\rho\sigma\tau}
 \overline{R}^{\nu}_{~\rho\sigma\tau}\right\}h_{\mu\nu}\nonumber\\ 
 &-&4\kappa\left\{\overline{R}^{\rho\nu}\overline{\nabla}_{\rho}\overline{\nabla}^{\mu} + \overline{R}^{\rho\mu}\overline{\nabla}_{\rho}\overline{\nabla}^{\nu} 
 - \overline{R}^{\mu\nu}\overline{\Box} + \overline{R}^{\rho\mu\sigma\nu}\overline{\nabla}_{\rho}\overline{\nabla}_{\sigma}\right\}h_{\mu\nu} 
 + 4\kappa \overline{R}^{\rho\sigma}\overline{\nabla}_{\rho}\overline{\nabla}_{\sigma}h^{\mu}_{\mu}
 \label{eft 4}
\end{eqnarray}
respectively. Therefore the conformal breaking Lagrangian (see Eq.(\ref{action part 3})) is expanded as,
\begin{eqnarray}
 \mathcal{L}_\mathrm{CB} = \sqrt{-\overline{g}}~\lambda\kappa^{2q}\left\{\overline{R}^q + \overline{\mathcal{G}}^{q/2} + \frac{\kappa}{2}\left(\overline{R}^q + 
 \overline{\mathcal{G}}^{q/2}\right)h^{\mu}_{\mu} + q\left(\frac{\delta R}{\overline{R}^{1-q}} + \frac{\delta \mathcal{G}}{2\overline{\mathcal{G}}^{1-q/2}}\right)\right\}
 F_{\mu\nu}\tilde{F}^{\mu\nu}~~,
 \label{eft 5}
\end{eqnarray}
where the overbar with a quantity indicates the respective quantity formed by the FRW metric $\overline{g}_{\mu\nu}$. The first two terms in the above 
expression, i.e $\sim \lambda\kappa^{2q}\big(\overline{R}^q + \overline{\mathcal{G}}^{q/2}\big)F\widetilde{F}$, encode the backreaction of the gauge fields 
on the background dynamics, while the rest of the above expression forms the interaction part between $h_{\mu\nu}$ and $A_{\alpha}$, in particular, 
\begin{eqnarray}
 \frac{1}{\sqrt{-\overline{g}}}\mathcal{L}_\mathrm{int}\left[h_{\mu\nu},A_{\alpha}\right] = 
 \lambda\kappa^{2q}\left\{\frac{\kappa}{2}\left(\overline{R}^q + 
 \overline{\mathcal{G}}^{q/2}\right)h^{\mu}_{\mu} + q\left(\frac{\delta R}{\overline{R}^{1-q}} + 
 \frac{\delta \mathcal{G}}{2\overline{\mathcal{G}}^{1-q/2}}\right)\right\}F_{\alpha\beta}\widetilde{F}^{\alpha\beta}~~.
 \label{eft 6}
\end{eqnarray}
It may be observed from Eq.(\ref{eft 6}) that the interaction Lagrangian acquires dimension 5 operators 
(like $F\tilde{F}h$) and dimension 7 operators (like $F\tilde{F}\partial\partial h$); in particular, we individually express such dimension 5 
(symbolized by $\mathcal{O}_{5}$) and dimension 7 ($\mathcal{O}_{7}$) interaction operators as follows,
\begin{eqnarray}
 \mathcal{O}_{5}&=&\lambda q~\frac{\kappa^{1+2q}}{\overline{R}^{1-q}}\bigg(-\overline{R}^{\mu\nu}\bigg)h_{\mu\nu}F_{\alpha\beta}\widetilde{F}^{\alpha\beta} 
 + \lambda q~\frac{\kappa^{1+2q}}{\overline{\mathcal{G}}^{1-q/2}}\bigg\{-\overline{R}^{\mu\nu}\overline{R} 
 + 2\left(4\overline{R}^{\rho\sigma}\overline{R}^{\mu~\nu}_{~\rho~\sigma} - \overline{R}^{\mu\rho\sigma\tau}
 \overline{R}^{\nu}_{~\rho\sigma\tau}\right)\bigg\}h_{\mu\nu}F_{\alpha\beta}\widetilde{F}^{\alpha\beta}\nonumber\\
 &+&\bigg(\frac{\lambda}{2}\bigg)\kappa^{1+2q}\bigg\{\overline{R}^{q} + \overline{\mathcal{G}}^{q/2}\bigg\}h^{\mu}_{\mu}
 F_{\alpha\beta}\widetilde{F}^{\alpha\beta}~~,
 \label{eft 7}
\end{eqnarray}
and
\begin{eqnarray}
 \mathcal{O}_{7}&=&\lambda q~\frac{\kappa^{1+2q}}{\overline{R}^{1-q}}\bigg\{\overline{\nabla}^{\mu}\overline{\nabla}^{\nu}h_{\mu\nu} 
 - \overline{\Box}h^{\mu}_{\mu}\bigg\}
 F_{\alpha\beta}\widetilde{F}^{\alpha\beta} + \lambda q~\frac{\kappa^{1+2q}}{\overline{\mathcal{G}}^{1-q/2}}\bigg\{\overline{R}\left(\overline{\nabla}^{\mu}
 \overline{\nabla}^{\nu}h_{\mu\nu} 
 - \overline{\Box}h^{\mu}_{\mu}\right)\nonumber\\ 
 &-&4\left(\overline{R}^{\rho\nu}\overline{\nabla}_{\rho}\overline{\nabla}^{\mu} + \overline{R}^{\rho\mu}\overline{\nabla}_{\rho}\overline{\nabla}^{\nu} 
 - \overline{R}^{\mu\nu}\overline{\Box} + \overline{R}^{\rho\mu\sigma\nu}\overline{\nabla}_{\rho}\overline{\nabla}_{\sigma}\right)h_{\mu\nu} 
 + 4 \overline{R}^{\rho\sigma}\overline{\nabla}_{\rho}\overline{\nabla}_{\sigma}h^{\mu}_{\mu}\bigg\}F_{\alpha\beta}\widetilde{F}^{\alpha\beta}~~,
 \label{eft 8}
\end{eqnarray}
respectively. Eq.(\ref{eft 7}) and Eq.(\ref{eft 8}) immediately argue that the dimension 5 and dimension 7 operators come with the following 
interaction coefficients,
\begin{eqnarray}
 \mathcal{C}_{(5)} \sim \lambda q~\kappa^{1+2q}\overline{R}^{q} \approx \lambda q~\kappa^{1+2q}\overline{\mathcal{G}}^{q/2}~~~~~~~~~\mathrm{and}~~~~~~~~~
 \mathcal{C}_{(7)} \sim \lambda q\bigg(\frac{\kappa^{1+2q}}{\overline{R}^{1-q}}\bigg) \approx \lambda q\bigg(\frac{\kappa^{1+2q}\overline{R}}
 {\overline{\mathcal{G}}^{1-q/2}}\bigg)~~,
 \label{eft 9}
\end{eqnarray}
which have mass dimension [-1] and [-3] respectively, as expected. We now estimate the cut-off the present magnetogenesis model by power 
counting of the operators present in the expression of the interaction Lagrangian \cite{Burgess:2009ea,Hertzberg:2010dc,Bezrukov:2010jz}. 
In particular, the presence of the dimension 
5 interaction operators introduce the cut-off scale ($\Lambda_{(5)}$) which can be estimated by,
\begin{eqnarray}
 \Lambda_{(5)} = \big[\mathcal{C}_{(5)}\big]^{\frac{-1}{5-4}} = \big[\mathcal{C}_{(5)}\big]^{-1} = 
 M_{\mathrm{Pl}}\bigg[\frac{1}{12^q\lambda q\big(H/M_\mathrm{Pl}\big)^{2q}}\bigg]~~,
 \label{eft 10}
\end{eqnarray}
where we use Eq.(\ref{eft 9}), and recall, $\kappa = \frac{1}{M_\mathrm{Pl}}$ and $H$ is the Hubble parameter during inflation. 
Similarly the cut-off introduced by the $\mathcal{O}_7$, is given by,
\begin{eqnarray}
 \Lambda_{(7)} = \big[\mathcal{C}_{(7)}\big]^{\frac{-1}{7-4}} = \big[\mathcal{C}_{(7)}\big]^{-1/3} 
 = M_\mathrm{Pl}\bigg[\frac{12^{2-2q}}{\lambda q}\bigg(\frac{H}{M_\mathrm{Pl}}\bigg)^{2-2q}\bigg]^{1/3}~~.
 \label{eft 11}
\end{eqnarray}
Clearly $\Lambda_{(7)} < \Lambda_{(5)}$, as $H \ll M_\mathrm{Pl}$ and also $\Lambda_{(7)}$ is suppressed by the exponent $1/3$. Thereby we may argue that 
the cut-off scale of the present model is given by,
\begin{eqnarray}
 \Lambda = \mathrm{min}\big[\Lambda_{(5)},\Lambda_{(7)}\big] = \Lambda_{(7)}~~,
 \label{eft 12}
\end{eqnarray}
that is obtained in Eq.(\ref{eft 11}). Having obtained the cut-off scale, we now investigate whether the relevant energy scale of the proposed model 
lies below than the cut-off. During the inflationary stage the typical momentum of the relevant excitations 
is equal to the Hubble parameter. Thus we determine the ratio $H/\Lambda$, in order to examine the validity of the present theory as an effective 
field theory, as follows,
\begin{eqnarray}
 \frac{H}{\Lambda} = \bigg[\frac{12^{2-2q}}{\lambda q}\bigg(\frac{M_\mathrm{Pl}}{H}\bigg)^{1+2q}\bigg]^{-1/3}~~.
 \label{eft 13}
\end{eqnarray}
As we have mentioned earlier that the present magnetogenesis scenario predicts sufficient magnetic strength at current universe when the 
parameter $q$ lies within $2.1 \leq q \leq 2.25$. With this information, we give the plots of $H/\Lambda$ with respect to $q$ in the range 
$2.1 \leq q \leq 2.25$, see Fig.[\ref{plot-eft}]. The blue curve and yellow curve represent the respective $H/\Lambda$ at 
the beginning of inflation (when $H = H_0 = 10^{13}\mathrm{GeV}$) and 
at the end of inflation (when $H = H_0\exp{\left[-\left(\frac{\delta}{\delta+1}\right) N_\mathrm{f}\right]}$, with 
$N_\mathrm{f}$ being the inflationary e-folding number) respectively. In the Fig.[\ref{plot-eft}], we take $N_\mathrm{f} = 51$. 
Fig.[\ref{plot-eft}] clearly demonstrates that the ratio $H/\Lambda$ during the inflation 
remains less than unity for the aforementioned range of $q$ which also leads to the correct magnetic field over the large scale modes at 
present epoch of the universe. The following points can be 
further argued from Fig.[\ref{plot-eft}]-- 
(a) $H/\Lambda$ at the end of inflation gets a lower value compared to that of at the beginning of inflation, 
and (b) the quantity $H/\Lambda$ seems to decrease as the value of $q$ 
increases. The fact that $H/\Lambda$ remains less than unity, i.e the relevant energy scale of the present model lies 
well below the cut-off scale, argues the validity of the proposed theory as an effective field theory. Therefore the regime 
of the parameter $q$, that makes the model viable in regard 
to the CMB observations of current magnetic strength and also makes the relevant energy scale of the model 
below than the cut-off scale, is given by $2.1 \leq q \leq 2.25$.
 
\section{Constraint from perturbative requirement}\label{sec-perturbative}

In this section, we derive a bound on the parameter space of the conformal breaking coupling function $f(R,\mathcal{G})$ such that the theory 
can be treated perturbatively, and the perturbative QFT makes sense. If we expand the metric as 
$g_{\mu\nu} = \overline{g}_{\mu\nu} + \kappa h_{\mu\nu}$, where $\overline{g}_{\mu\nu}$ is the background FRW metric and $h_{\mu\nu}$ are 
the metric perturbations, then the conformal breaking action $S_\mathrm{CB}$ of Eq.(\ref{action part 3}) introduces non-minimal 
interaction terms between the graviton and photon. Such interaction Lagrangian is obtained in Eq.(\ref{eft 6}) as,
\begin{eqnarray}
 \frac{1}{\sqrt{-\overline{g}}}\mathcal{L}_\mathrm{int}\left[h_{\mu\nu},A_{\alpha}\right] = 
 \lambda\kappa^{2q}\left\{\frac{\kappa}{2}\left(\overline{R}^q + 
 \overline{\mathcal{G}}^{q/2}\right)h^{\mu}_{\mu} + q\left(\frac{\delta R}{\overline{R}^{1-q}} + \frac{\delta \mathcal{G}}{2\overline{\mathcal{G}}^{1-q/2}}\right)\right\}
 F_{\alpha\beta}\widetilde{F}^{\alpha\beta}~~,
 \label{per 1}
\end{eqnarray}
where $\delta R$ and $\delta \mathcal{G}$ are obtained in Eq.(\ref{eft 3}) and Eq.(\ref{eft 4}) respectively. The above interaction terms 
contribute in the Feynman-Dyson series of the 2-point correlator of EM field, and from the perturbative requirement, we demand that the first 
terms in the Feynman-Dyson series to be small. In particular, the constraint on the coupling function from perturbative requirement 
can be derived by either of the following two conditions:
\begin{enumerate}
 \item the ratio of the actions for the conformal breaking term to the canonical electromagnetic term should be less than unity \cite{Durrer:2010mq}, i.e,
 \begin{eqnarray}
  \left|\frac{S_\mathrm{CB}}{S_\mathrm{em}^{(can)}}\right| < 1~~.
  \label{perturbative condition 1}
 \end{eqnarray}
 
 \item The loop contribution in the EM field propagator should be less than that of the tree propagator \cite{Ferreira:2015omg}. 
 In particular,
 \begin{eqnarray}
  \left|\frac{\big\langle AA \big\rangle_{1-loop}}{\big\langle AA \big\rangle_{tree}}\right| < 1~~,
  \label{perturbative condition 2}
 \end{eqnarray}
where $\big\langle AA \big\rangle_{tree}$ represents the tree propagator of the EM field and 
$\big\langle AA \big\rangle_{1-loop}$ indicates the loop correction in the EM 2-point correlator. 
\end{enumerate}
Here we would like to mention that these two conditions are 
equivalent, as the loop contribution in the EM propagator arises due to the presence 
of the action $S_\mathrm{CB}$.

\begin{figure}[!h]
\begin{center}
 \centering
 \includegraphics[width=3.0in,height=2.5in]{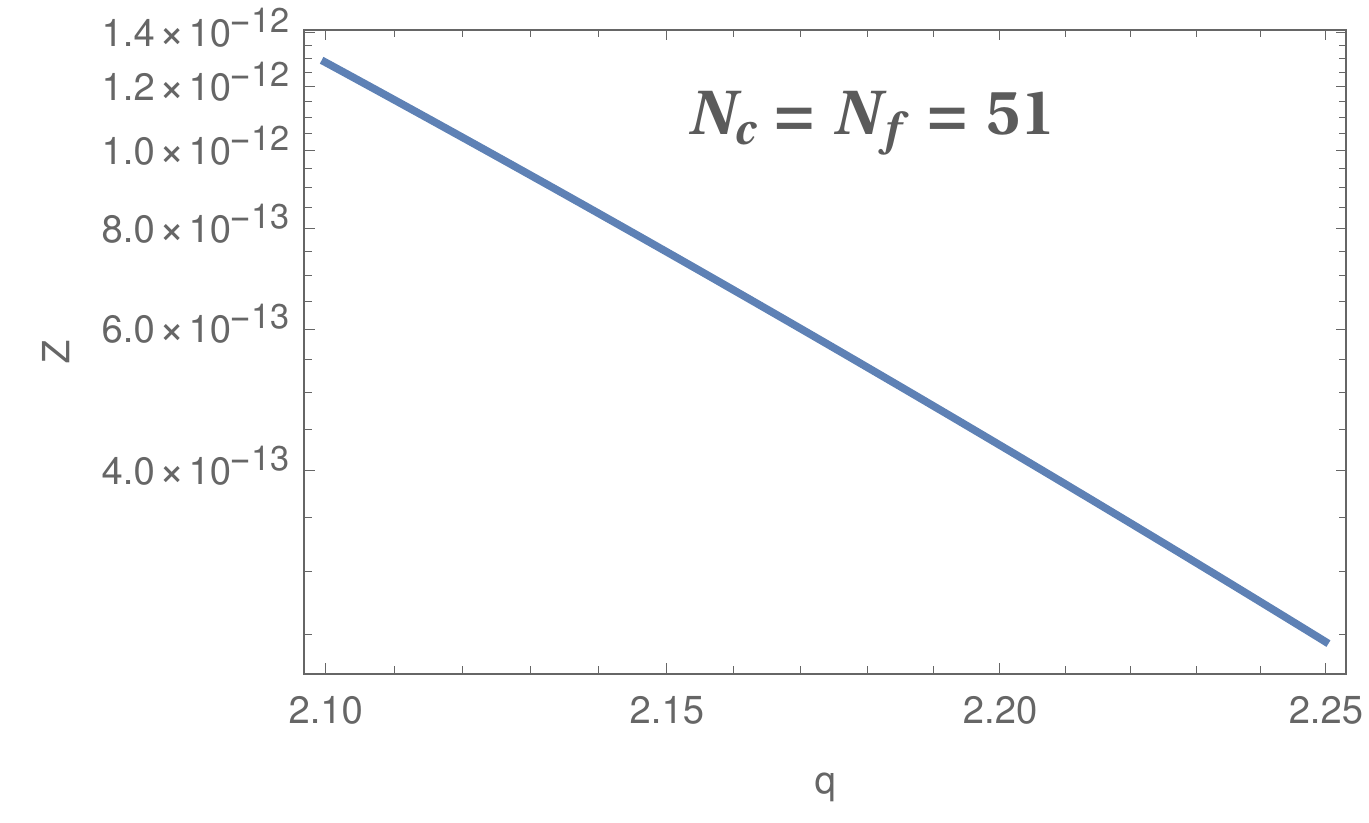}
 \includegraphics[width=3.0in,height=2.5in]{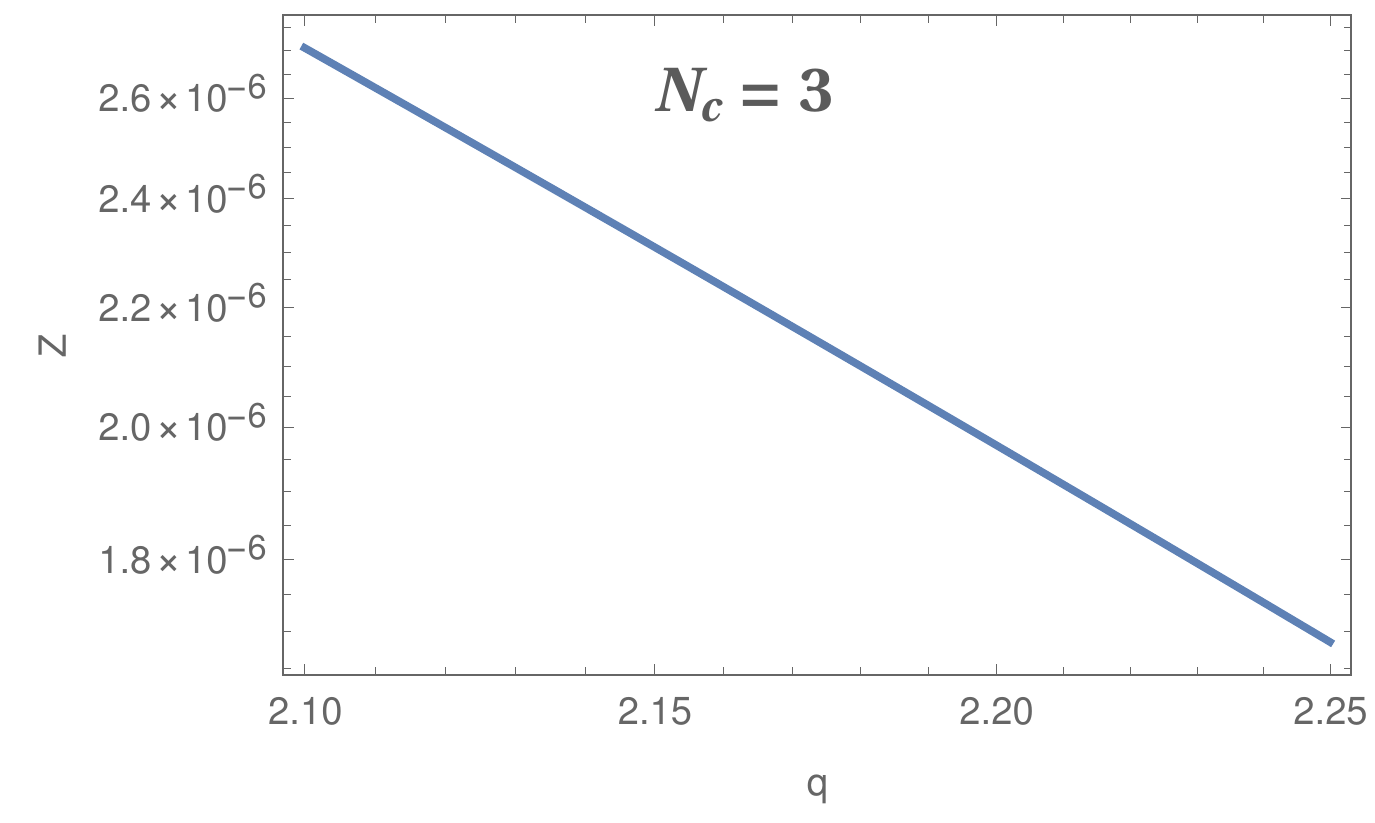}
 \caption{$\mathcal{Z}$ vs $q$ for $H_0 = 10^{13}\mathrm{GeV}$, $N_\mathrm{f} = 51$ and $\delta = 0.1$. 
 Moreover we take $\lambda = 1$. In the left plot, 
 $k_c = 10^{-20}\mathrm{GeV}$ that crosses the horizon near the end of inflation, i.e $N_\mathrm{c} \approx N_\mathrm{f} = 51$; while for the right plot, 
 $k_c = 10^{-38}\mathrm{GeV}$ which crosses the horizon near $N_c \approx 3$ i.e near the beginning of inflation.}
 \label{perturbative}
\end{center}
\end{figure}

To examine the first condition in the present context, we start with the following expression of the canonical EM Lagrangian,
\begin{eqnarray}
 -\frac{1}{4}F_{\mu\nu}F^{\mu\nu} = \frac{1}{a^4}\left(-\frac{1}{2}\left(A_i'\right)^2 + \frac{1}{4}F_{ij}F_{ij}\right) = \rho(\vec{E}) - \rho(\vec{B})~~,
 \label{perturbative 1}
\end{eqnarray}
where $\rho(\vec{E})$ and $\rho(\vec{B})$ are the electric and the magnetic energy density respectively. Consequently, 
the canonical EM action takes the following form,
\begin{eqnarray}
 S_\mathrm{em}^{(can)}&=&\int d\eta d^3x~a^4\big(\rho(\vec{E}) - \rho(\vec{B})\big)\nonumber\\
 &=&V\int d\eta~a^4\bigg(\big\langle \rho(\vec{E}) \big\rangle_{V} - \big\langle \rho(\vec{B}) \big\rangle_{V}\bigg)~~,
 \label{perturbative 2}
\end{eqnarray}
with $\big\langle...\big\rangle_{V}$ denotes the average over a spatial volume $V$ and is considered to be equivalent 
to the vacuum expectation value over the Bunch-Davies state (defined in Eq.(\ref{electric power spectrum 1}) or in Eq.(\ref{magnetic power spectrum 1})). 
In particular,
\begin{eqnarray}
 a^4\big\langle \rho(\vec{E}) \big\rangle_{V}&=&\sum_{r=1,2}\int \frac{k^2}{2\pi^2}\big|A_r'(k,\eta)\big|^2~dk~~,\nonumber\\
 a^4\big\langle \rho(\vec{B}) \big\rangle_{V}&=&\sum_{r=1,2}\int \frac{k^4}{2\pi^2}\big|A_r(k,\eta)\big|^2~dk~~.
 \label{perturbative 3}
 \end{eqnarray}
For the purpose of determining the $S_\mathrm{CB}$, we express $\epsilon^{\mu\nu\alpha\beta}F_{\mu\nu}F_{\alpha\beta} = F\widetilde{F}$, in the 
language of differential forms, as,
\begin{eqnarray}
 \sqrt{-g}F\widetilde{F}d^4x = 4F \wedge F~~~~~~~~~~~~~~\mathrm{and}~~~~~~~~~~~~~~~F = dA~~.
 \label{perturbative 4}
\end{eqnarray}
Therefore the conformal breaking action turns out to be,
\begin{eqnarray}
 S_\mathrm{CB} = 4\int \lambda f(R,\mathcal{G}) \big(dA \wedge F\big) = 4\int \lambda\big(df \wedge F \wedge A\big)~~.
 \label{perturbative 5}
\end{eqnarray}
To arrive at the second equality of the above expression, we use the integration by parts. Considering the comoving observer 
(having four velocity $u^{\mu} = \left(a^{-1},0,0,0\right)$ or $u = -ad\eta$) for measuring the electric and magnetic fields, we find 
$df \wedge F \wedge A = 2f'(\eta)a^3\rho_h$ (with $\rho_h$ being the helicity density) \cite{Durrer:2010mq}. Accordingly the 
$S_\mathrm{CB}$ becomes,
\begin{eqnarray}
 S_\mathrm{CB} = 8\int d\eta d^3x~\lambda f'(\eta)\big(a^3\rho_h\big) = 8\int d\eta~\lambda f'(\eta)~a^3\big\langle \rho_h \big\rangle_{V}~~,
 \label{perturbative 6}
\end{eqnarray}
where $\big\langle \rho_h \big\rangle_{V}$ is given by,
\begin{eqnarray}
 a^3\big\langle \rho_h \big\rangle_{V} = \int \frac{k^3}{2\pi^2}\left\{\big|A_{+}(k,\eta)\big|^2 - \big|A_{-}(k,\eta)\big|^2\right\}dk~~.
 \label{perturbative 7}
\end{eqnarray}
Plugging back the above expressions into the left hand side of Eq.(\ref{perturbative condition 1}), we arrive at the following equation,
\begin{eqnarray}
 \left|\frac{S_\mathrm{CB}}{S_\mathrm{em}^{(can)}}\right|&=&
 \left|\frac{8\int d\eta~\lambda f'(\eta)~a^3\big\langle \rho_h \big\rangle_{V}}
 {\int d\eta~a^4\bigg(\big\langle \rho(\vec{E}) \big\rangle_{V} - \big\langle \rho(\vec{B}) \big\rangle_{V}\bigg)}\right|~.
 \label{perturbative 11}
\end{eqnarray} 
Now, for the condition $\left|\frac{S_\mathrm{CB}}{S_\mathrm{em}^{(can)}}\right| < 1$ to be satisfied, it is sufficient to 
require
\begin{eqnarray}
 \frac{8\lambda f'(\eta)~a^3\big\langle \rho_h \big\rangle_{V}}
 {a^4\left(\big\langle \rho(\vec{E}) \big\rangle_{V} - \big\langle \rho(\vec{B}) \big\rangle_{V}\right)} < 1~~.
 \label{perturbative 12}
\end{eqnarray}
Let us denote the ratio in the left hand side of Eq.(\ref{perturbative 12}) by $\mathcal{Z}$. Eq.(\ref{form of f 2}) immediately leads to $f'(\eta)$ as,
\begin{eqnarray}
 f'(\eta) = \frac{1}{8\lambda}\left(\frac{\zeta^2}{\eta^2}\right)\left(\frac{-\eta_0}{\eta}\right)^{2\alpha}~~,
 \label{perturbative 13}
\end{eqnarray}
where $\zeta^2$ is given in Eq.(\ref{B}), due to which, $\mathcal{Z}$ can be equivalently expressed as,
\begin{eqnarray}
 \mathcal{Z} = H_0\left(16\epsilon q\lambda\right)\left(\frac{H_0}{M_\mathrm{Pl}}\right)^{2q}\left\{\big[6\beta(\beta+1)\big]^q 
 + \big[-24(\beta+1)^3\big]^{q/2}\right\}\left(\frac{-\eta_0}{\eta}\right)^{1+2\alpha}
 \left[\frac{\big\langle \rho_h \big\rangle_{V}}
 {\left(\big\langle \rho(\vec{E}) \big\rangle_{V} - \big\langle \rho(\vec{B}) \big\rangle_{V}\right)}\right]~~.
 \label{perturbative 14}
\end{eqnarray}
Moreover from Eq.(\ref{electric power spectrum 1}), Eq.(\ref{magnetic power spectrum 1}) and Eq.(\ref{helicity power spectrum inflation}), 
we have the following expressions,
\begin{eqnarray}
 \big\langle \rho(\vec{E},\eta_c) \big\rangle_{V}&=&
 \int_{k_i}^{k_c} \frac{dk}{2\pi^4}\left(\frac{H_0}{k}\right)^2H^{4+4\delta}\left(-k\eta\right)^{4+4\delta}
 \left|\left(\frac{\zeta\sqrt{k}}{2\alpha}\right)^{-\frac{1}{2\alpha}}\Gamma\left(\frac{1}{2\alpha}\right)\right|^2
 \left\{\left|C_2\right|^2\right\}~~,\nonumber\\
 \big\langle \rho(\vec{B},\eta_c) \big\rangle_{V}&=&
 \int_{k_i}^{k_c} \frac{dk}{2\pi^2}H^{4+\delta}\left(-k\eta\right)^{4+\delta}\left|\frac{\left(\frac{\zeta\sqrt{k}}{2\alpha}\right)^{\frac{1}{2\alpha}}}
 {\Gamma\left(1 + \frac{1}{2\alpha}\right)}\right|^2
 \left\{\left|C_1 - C_2\cot{\left(-\frac{\pi}{2\alpha}\right)}\right|^2\right\}~~,\nonumber\\
 \big\langle \rho_h(\eta_c) \big\rangle_{V}&=&
 \int_{k_i}^{k_c} \frac{dk}{2\pi^2}H^{3+3\delta}\left(-k\eta\right)^{3+3\delta}\left|\frac{\left(\frac{\zeta\sqrt{k}}{2\alpha}\right)^{1/(2\alpha)}}
 {\Gamma\left(1 + \frac{1}{2\alpha}\right)}\right|^2
 \left\{\left|C_1 - C_2\cot{\left(-\frac{\pi}{2\alpha}\right)}\right|^2\right\}~~,
 \label{perturbative 15}
\end{eqnarray}
respectively, where $C_i$ ($i=1,2,3,4$) are shown in the Appendix. The integration limit in Eq.(\ref{perturbative 15}) is taken from 
$k_i$ to $k_c$, i.e from the mode that crosses the horizon at the beginning of inflation to the mode which crosses the horizon 
at the instance $\eta = \eta_c$. Now we identify the beginning of inflation when the horizon is of same size 
with the CMB scale mode, i.e we may write $k_i = k_{CMB} = 0.05\mathrm{Mpc}^{-1}$. Furthermore we have $|k_c\eta_c| = 1$ with 
$\eta_c$ is any time during the inflation and thus $k_c > k_{CMB}$. The quantity $N_\mathrm{c}$ is the 
e-folding number up-to $\eta=\eta_c$ measured from the beginning of inflation, 
i.e $N_\mathrm{c} = \ln{\left(\frac{a_c}{a_{beg}}\right)}$ with $a_c = a(\eta_c)$ and $a_{beg} = a(\eta_{beg})$. 
Having obtained the necessary ingredients, we now examine whether the condition $\mathcal{Z} < 1$ is satisfied during inflation. 
However due to the dependence of $C_i = C_i(k)$ ($i=1,2,3,4$), the integrations in Eq.(\ref{perturbative 15}) 
may not be obtained in analytic form(s), and thus we numerically approach to integrate 
$\big\langle \rho(\vec{E}) \big\rangle_{V}$, $\big\langle \rho(\vec{B}) \big\rangle_{V}$ and 
$\big\langle \rho_h \big\rangle_{V}$ (at $\eta_c$) present in Eq.(\ref{perturbative 15}). 
For this purpose, we consider $H_0 = 10^{13}\mathrm{GeV}$, $N_\mathrm{f} = 51$ and $\delta =  0.1$ respectively, and 
perform the numerical integrations of Eq.(\ref{perturbative 15}). Consequently we depict the 
plot of $\mathcal{Z}$ with respect to the parameter $q$ in the range $2.1 \leq q \leq 2.25$, see Fig.[\ref{perturbative}]. 
Recall this range of $q$ results to 
the correct magnetic strength at present epoch of the universe, and thus we are using such range of $q$ to examine the perturbative condition 
in order to keep intact the generation of the EM field. 
We consider different values of $k_c$ in Fig.[\ref{perturbative}], in particular, 
we consider $k_c = 10^{-20}\mathrm{GeV}$ and $10^{-38}\mathrm{GeV}$ in the left and 
right plot of Fig.[\ref{perturbative}] respectively. Here we would like to mention that the mode 
$k_c = 10^{-20}\mathrm{GeV}$ crosses the horizon near the end of inflation, i.e $N_c \approx N_f$; while the mode $k_c = 10^{-38}\mathrm{GeV}$ crosses 
the horizon near $N_c \approx 3$ i.e near the beginning of inflation. 

The Fig.[\ref{perturbative}] clearly demonstrates that the perturbative condition $\mathcal{Z} < 1$ is satisfied for 
$q = [2.1,2.25]$ which also leads to the viability of the model in regard to the CMB observations of current magnetic strength. 
Therefore the predictions of perturbative QFT in the model are found to be consistent with the observational bound of the model parameter required to get 
sufficient magnetic strength at current stage of the universe.
 
 \section{Curvature perturbation sourced by electromagnetic field during inflation}\label{sec-perturbation}
 
The produced electromagnetic field during inflation may induce the curvature perturbation 
\cite{Fujita:2013qxa,Barnaby:2012tk,Bamba:2014vda,Suyama:2012wh}, and the power spectrum of the induced curvature 
perturbations should satisfy the recent Planck constraints. Thereby in the present magnetogenesis scenario where the electromagnetic field couples 
with the background curvature terms via the dual field tensor, it is important to examine the viability of the sourced curvature perturbations 
in respect to the Planck constraints.

The induced curvature perturbation (symbolized by $\chi(\eta,\vec{x})$) from the electromagnetic field is expressed as \cite{Fujita:2013qxa},
\begin{eqnarray}
 \chi_{em}(\eta,\vec{x}) = -\frac{2H}{\epsilon\rho_{inf}}\int_{\eta_m}^{\eta} d\eta'~a(\eta') \rho_{em}(\eta',\vec{x})
 \label{induced1}
\end{eqnarray}
where $\rho_{inf}$ is the background inflaton energy density and $\rho_{em}$ denotes the EM field 
energy density. Here it may be mentioned that the contribution from the 
electromagnetic anisotropic stress is suppressed compared to the contribution written in the r.h.s of Eq.(\ref{induced1}) (see \cite{Suyama:2012wh}), 
and thus the electromagnetic anisotropic stress in the curvature perturbation is not taken into account in Eq.(\ref{induced1}). 
The lower limit of the integral, i.e $\eta_m$, represents the time at which the EM production effectively starts.

\begin{figure}[!h]
\begin{center}
 \centering
 \includegraphics[width=3.5in,height=2.5in]{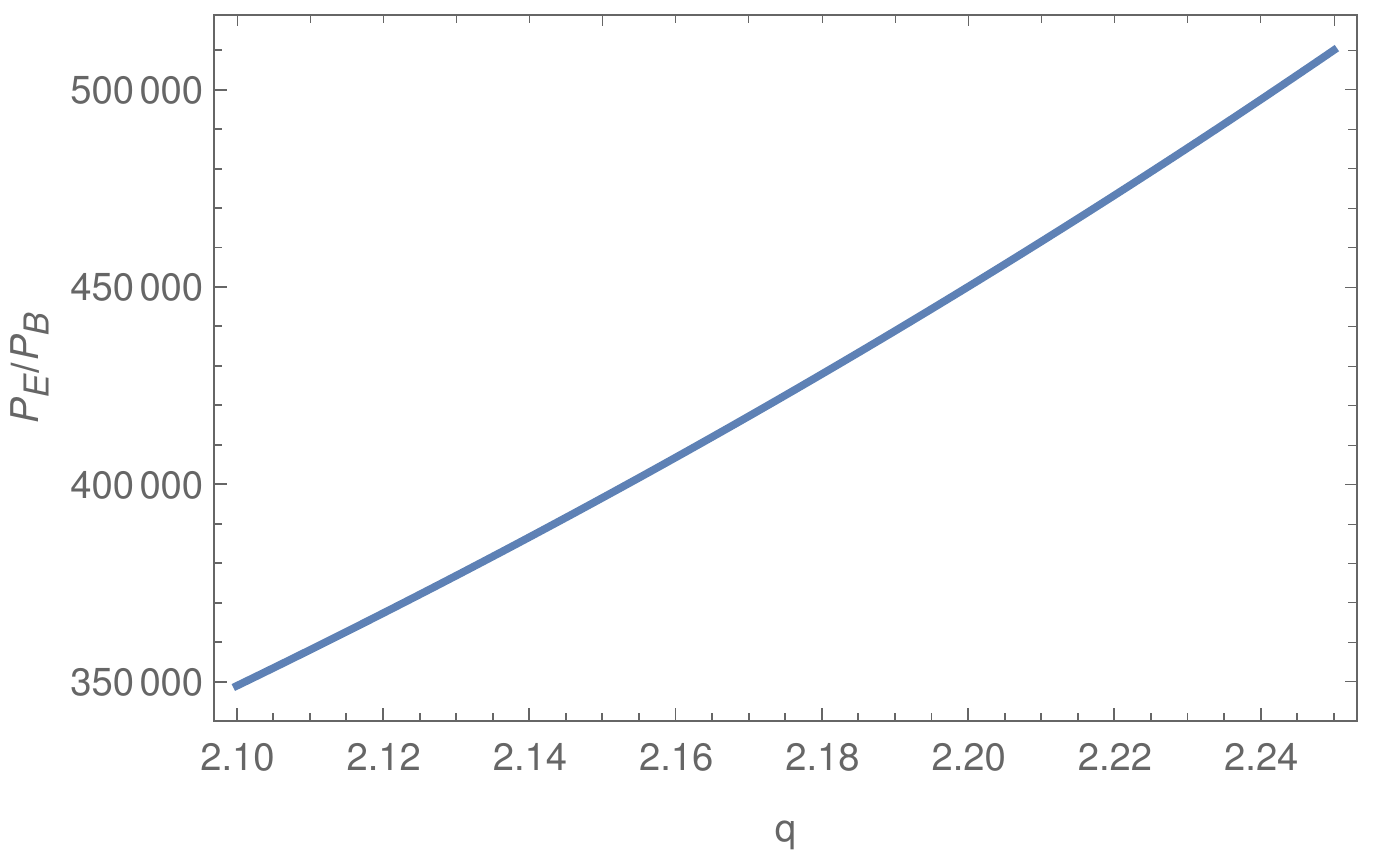}
 \caption{$\frac{\mathcal{P}(\vec{E})}{\mathcal{P}(\vec{B})}$ vs $q$ in the range $2.1 \leq q \leq 2.25$. Here we consider 
 $H_0 = 10^{13}\mathrm{GeV}$, $N_\mathrm{f} = 51$ and $\delta = 0.1$. 
 Moreover we take $\lambda = 1$.}
 \label{plot-ratio-EB}
\end{center}
\end{figure}

The EM energy density can be expressed by $\rho_{em}(\eta,\vec{x}) = \rho_E(\eta,\vec{x}) + \rho_B(\eta,\vec{x})$, where $\rho_E(\eta,\vec{x})$ and 
$\rho_B(\eta,\vec{x})$ are the energy density for electric and magnetic fields respectively. However from Eq.(\ref{electric power spectrum 1}) 
and Eq.(\ref{magnetic power spectrum 1}), the ratio of electric to magnetic power spectrum during inflation 
comes as $\frac{\mathcal{P}(\vec{E})}{\mathcal{P}(\vec{B})} \sim 10^{5}$. 
In particular, we give the plot of $\frac{\mathcal{P}(\vec{E})}{\mathcal{P}(\vec{B})}$ with respect to $q$ in the range 
$2.1 \leq q \leq 2.25$ on which we are interested, see Fig.[\ref{plot-ratio-EB}]. 

The figure clearly depicts that the electric field 
during inflation is $\sim 10^{5}$ times stronger than that of the magnetic field strength. This in turn indicates that 
the main contribution of the EM energy density comes from the electric field, and thus we may write 
$\rho_{em} \approx \rho_E = \frac{1}{2}E^2$. Consequently the EM field energy density in Fourier space is given by,
\begin{eqnarray}
 \rho_{em}(\eta_\mathrm{f},\vec{k}) = \frac{1}{2}\int\int \frac{d^3p_1d^3p_2}{(2\pi)^3}~\delta\big(\vec{p}_1 + \vec{p}_2 - \vec{k}\big) 
 \vec{E}(\eta_\mathrm{f},\vec{p}_1)\vec{E}(\eta_\mathrm{f},\vec{p}_2)~~,
 \label{induced2}
\end{eqnarray}
where the electric field is defined as $\big|E(\eta,k)\big| = \frac{1}{a^2}\big|A'(k,\eta)\big|$ with respect to the comoving observer. 
Thereby Eq.(\ref{superhorizon form 2}) immediately leads to the electric field as,
\begin{eqnarray}
 \big|E(\eta,k)\big| = 
 k\left(\frac{H_0}{k}\right)\left|\left(\frac{C_2\Gamma\left(\frac{1}{2\alpha}\right)}{\pi}\right)
 \left(\frac{\zeta\sqrt{k}}{2\alpha}\right)^{-1/(2\alpha)}\right|\left(-H\eta\right)^2~~.
 \label{induced3}
\end{eqnarray}
With the above expression of $\big|E(\eta,k)\big|$, we evaluate the 2-point correlator of $\zeta_{em}(\eta,\vec{k})$ 
in the present context as \cite{Fujita:2013qxa},
\begin{eqnarray}
 \langle\chi_{em}(\eta_f,\vec{k}_1)\chi_{em}(\eta_f,\vec{k}_2)\rangle&=&2\delta\big(\vec{k}_1 + \vec{k}_2\big) G_2 
 \int_{k_{CMB}}^{k_{f}} d^3p_1d^3p_2 \delta\left(\vec{p}_2 - \vec{p}_1 - \vec{k}_2\right) \left[f(p_1)f(p_2)\right]^2\nonumber\\
 &\times&\left(\delta_{j_1j_2} - \big(\hat{p}_1\big)_{j_1}\big(\hat{p}_1\big)_{j_2}\right)
 \left(\delta_{j_1j_2} - \big(\hat{p}_2\big)_{j_1}\big(\hat{p}_2\big)_{j_2}\right)
 \bigg\{\prod_{i=1,2}~\int_{\eta_m}^{\eta_f}d\eta_i\big(-\eta_i\big)^{3+3\delta}\bigg\}
 \label{induced4}
\end{eqnarray}
where $G_2$ and $f(p)$ have the following forms,
\begin{eqnarray}
 G_2 = \bigg[\frac{H_f^2}{6\epsilon M_\mathrm{Pl}^2}\bigg]^2~~,
 \label{G}
\end{eqnarray}
and
\begin{eqnarray}
 f(k) = k\left(\frac{H_0}{k}\right)\left|\left(\frac{C_2\Gamma\left(\frac{1}{2\alpha}\right)}{\pi}\right)
 \left(\frac{\zeta\sqrt{k}}{2\alpha}\right)^{-1/(2\alpha)}\right|~~,
 \label{f}
\end{eqnarray}
respectively. Here $k_\mathrm{f}$ in Eq.(\ref{induced4}) symbolizes the mode that crosses the horizon at the end of inflation. Moreover, 
to derive $G_2$, we use $\rho_{inf}(\eta_f) = 3H_f^2M_\mathrm{Pl}^2$. Such expression of $\rho_{inf}$ holds true as the EM field provides 
a negligible backreaction on the background spacetime in the present magnetogenesis scenario. 
We may perform the $p_2$ integral of Eq.(\ref{induced4}), to get
\begin{eqnarray}
 \langle\chi_{em}(\eta_f,\vec{k}_1)\chi_{em}(\eta_f,\vec{k}_2)\rangle = \frac{32\pi}{3}\delta\big(\vec{k}_1 + \vec{k}_2\big) G_2 
 \int_{k_{CMB}}^{k_f} dp_1~p_1^{2}~\left[f(p_1)f(p_1 + k_2)\right]^2\bigg\{\prod_{i=1,2}~\int_{-1/p_1}^{\eta_f}d\eta_i\big(-\eta_i\big)^{3+3\delta}\bigg\}~.
 \label{induced5}
\end{eqnarray}
where we use the integral $\int d\Omega_k\hat{k}_i\hat{k}_j = \frac{4\pi}{3}\delta_{ij}$. For the momentum variable $p_1$ in the above integral, 
the corresponding lower limit of the $\eta$ integral is 
taken as 
\begin{eqnarray}
 \eta_m = -\frac{1}{p_1}~~,
 \label{neweq}
\end{eqnarray}
i.e when the mode $p_1$ crosses the horizon. This is due to the reason that the EM fluctuations of 
momentum $p_1$ starts to effectively produce from the horizon crossing of $p_1$. In particular, 
the energy density stored in a certain mode of the gauge field is maximal (compared to the 
background energy density) at horizon crossing of the corresponding mode and then redshifts almost like radiation. Therefore a certain EM mode 
is mainly produced near the horizon crossing of that mode in the present magnetogenesis scenario. The 
consideration of $\eta_m = -1/p_1$ indeed takes care the horizon crossing region of the mode variable $p_1$. With $\eta_m = -1/p_1$ and $\eta_f = -1/k_f$, 
we evaluate the $\eta$ integral of Eq.(\ref{induced5}), and get
\begin{eqnarray}
\langle\chi_{em}(\eta_f,\vec{k}_1)\chi_{em}(\eta_f,\vec{k}_2)\rangle = \frac{32\pi}{3}\delta\big(\vec{k}_1 + \vec{k}_2\big) G_2 
 \int_{k_{CMB}}^{k_f} dp_1~p_1^{2}~\left[f(p_1)f(p_1 + k_2)\right]^2
 \bigg\{\frac{\left(1/p_1\right)^{4+3\delta} - \left(1/k_f\right)^{4+3\delta}}{4+3\delta}\bigg\}^2~.
 \label{induced-new1}                                          
\end{eqnarray}                                    
We will eventually evaluate the two point correlator at CMB scale, 
and thus $k_1 = k_2 = k_{CMB}$. 
The above expression of 2-point correlator yields the power spectrum of the curvature perturbation (at $k = k_1$) induced by the EM field as,
\begin{eqnarray}
 \mathcal{P}(\chi_{em}, k_1) =  G_2\left(\frac{16}{3\pi}\right)k_1^3\int_{k_{1}}^{k_f} dp_1~p_1^{2}~\left[f(p_1)f(p_1 + k_2)\right]^2
 \bigg\{\frac{\left(1/p_1\right)^{4+3\delta} - \left(1/k_f\right)^{4+3\delta}}{4+3\delta}\bigg\}^2~~,
 \label{induced power spectrum}
\end{eqnarray}
where the functional form of $f(p_1)$ or $f(p_1+k_2)$ are shown in Eq.(\ref{f}).\\

Having obtained the theoretical expression of induced power spectrum in hand, we now confront the model with the Planck 
results which put constraint on curvature perturbation as,
\begin{eqnarray}
 \mathcal{P}^{obs}(\chi) \approx 2.1\times10^{-9}~~.
 \label{constraints}
\end{eqnarray}
We consider that the dominant component of the power spectrum of the curvature perturbation is generated by the background slow-roll inflaton field. 
As a consequence, the theoretical prediction of 
$\mathcal{P}(\chi_{em})$ does not exceed the aforementioned Planck constraint, in particular,
\begin{eqnarray}
 \mathcal{P}(\chi_{em}) < \mathcal{P}^{obs}(\chi)~~.
 \label{constraint Nm}
\end{eqnarray}

\begin{figure}[!h]
\begin{center}
 \centering
 \includegraphics[width=3.5in,height=2.5in]{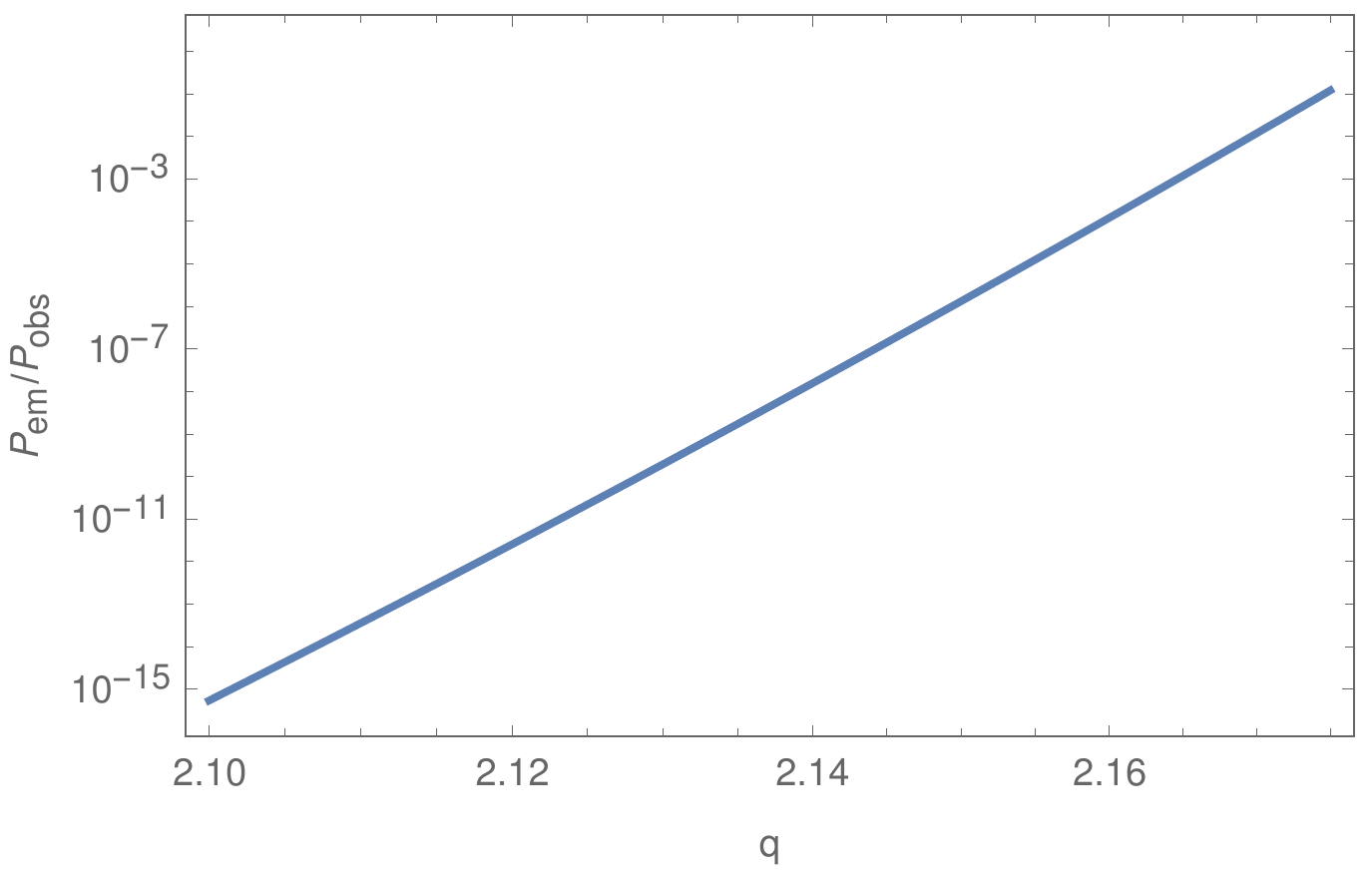}
 \caption{$\frac{\mathcal{P}(\chi_{em})}{\mathcal{P}^{obs}(\chi)}$ vs $q$. Here we consider 
 $H_0 = 10^{13}\mathrm{GeV}$, $N_f = 51$ and $\delta = 0.1$. Moreover we take $\lambda = 1$.}
 \label{numerics}
\end{center}
\end{figure}

In order to investigate $\mathcal{P}(\chi_{em}) < \mathcal{P}^{obs}(\chi)$ in the present context, 
we need to evaluate the $p_1$ integral of Eq.(\ref{induced power spectrum}). 
However due to the aforementioned form of $f(k)$, this integral 
may not be obtained in a closed form, so we perform the integration by numerical analysis. This is depicted in Fig.[\ref{numerics}] where 
we take the following set of parameters: $H_0 = 10^{13}\mathrm{GeV}$, $\delta = 0.1$, $N_\mathrm{f} = 51$ and $\lambda = 1$. 
In particular, we plot 
the ratio of $\frac{\mathcal{P}(\chi_{em})}{\mathcal{P}^{obs}(\chi)}$ with respect to the parameter $q$ in Fig.[\ref{numerics}].\\ 

The Fig.[\ref{numerics}] clearly demonstrates that in order to satisfy $\mathcal{P}(\chi_{em}) < \mathcal{P}^{obs}(\chi)$, the parameter $q$ should lie within 
$q \lesssim 2.175$. Moreover we recall that the magnetogenesis model under consideration predicts correct magnetic strength at present universe for 
$2.1 \lesssim q \lesssim 2.25$. Therefore it turns out that the whole range of $q$ which gives the correct magnetic strength, does not obey 
the condition of the induced curvature perturbation i.e $\mathcal{P}(\chi_{em}) < \mathcal{P}^{obs}(\chi)$. In particular, 
the range of $q$ which leads to a sufficient magnetic strength at present universe and also ensures 
$\mathcal{P}(\chi_{em}) < \mathcal{P}^{obs}(\chi)$, is given by: $2.1 \lesssim q \lesssim 2.175$.\\

Before concluding we would like to mention that some recent literatures have argued that 
non-linear enhancement of the magnetic fields at the end of inflation, inverse cascade of helical photons 
after inflation and/or a simultaneous coupling to the photon kinetic term $F_{\mu\nu}F^{\mu\nu}$ could help increase 
the strength of the magnetic field \cite{Adshead:2016iae,Fujita:2019pmi}. 
Such considerations in the present curvature coupled helical magnetogenesis scenario will be examined 
in future work.

\section{Conclusion}
The recently proposed curvature coupling helical magnetogenesis scenario \cite{Bamba:2021wyx}, 
where the EM field couples with the background $f(R,\mathcal{G})$ gravity, 
has the following strong features -- (1) the model predicts sufficient magnetic strength 
at current epoch of the universe for suitable range of the model parameter ($q$) given by: $2.1 \leq q \leq 2.26$ for instantaneous reheating scenario and 
$2.1 \leq q \leq 2.25$ for Kamionkowski reheating scenario respectively; (2) the EM field is found to have 
a negligible backreaction over the background spacetime and does not jeopardize the background inflation; (3) the model is free from the strong couping problem; 
(4) due to the helical nature of the magnetic field, it turns out that the magnetic strength of $\sim 10^{-13}\mathrm{G}$ over the galactic scale 
results to the correct baryon asymmetry of the universe consistent with the observational data.\\ 

However in the realm of 
inflationary magnetogenesis, the above requirements are not enough to argue about the viability of the model. In particular, one needs to examine 
some more important requirements to ensure the viability of a magnetogenesis model, such as -- (1) whether the model is consistent 
with the predictions of perturbative QFT, as the calculations that we use to determine the magnetic field's evolution and its power spectrum are based on the perturbative 
QFT; (2) the curvature perturbation sourced by the EM field during inflation should not exceed than the curvature perturbation 
contributed from the background inflaton field, in order to be consistent with the recent Planck data; and 
(3) the relevant energy scale of the magnetogenesis model needs to be lie below than 
the cut-off scale of the model. We have checked all these requirements in the present context of curvature coupling 
helical magnetogenesis scenario. For the perturbative requirement, we have examined 
whether the condition $\left|\frac{S_{CB}}{S_{can}}\right| < 1$ is satisfied, where $S_{can}$ and $S_{CB}$ are the 
canonical and the conformal breaking action of the EM field respectively. The condition 
$\left|\frac{S_{CB}}{S_{can}}\right| < 1$ actually indicates that the loop contribution of EM two-point correlator 
is less than the tree propagator of the EM field -- which is the essence of the perturbative quantum field theory. For the second requirement, we 
have calculated the power spectrum of curvature perturbation sourced by the EM field at super-Hubble scales ($\mathcal{P}(\chi_{em})$). 
By considering that the primordial 
curvature perturbation is mainly contributed from the slow-roll inflaton field, we have determined the necessary condition corresponding to the 
requirement given by: $\mathcal{P}(\chi_{em}) < \mathcal{P}^{obs}(\chi)$, where $\mathcal{P}^{obs}(\chi)$ corresponds to the Planck observation 
of the curvature perturbation power spectrum. This puts a constraint on the parameter $q$ as $q \lesssim 2.175$. 
Therefore it turns out that the whole range of $q$ which gives the correct magnetic strength, does not obey 
the condition of the induced curvature perturbation i.e $\mathcal{P}(\chi_{em}) < \mathcal{P}^{obs}(\chi)$. In particular, 
the range of $q$ which leads to a sufficient magnetic strength at present universe and also ensures 
$\mathcal{P}(\chi_{em}) < \mathcal{P}^{obs}(\chi)$, is given by: $2.1 \lesssim q \lesssim 2.175$.\\  

Interestingly, all the three aforementioned requirements are found to be simultaneously satisfied 
by that range of the model parameter which leads to the correct magnetic strength over the large scale modes.

\section{Appendix: Forms of $C_i$ ($i=1,2,3,4$)}\label{sec-app}

The solutions of $A_{\pm}(k,\eta)$ can be demonstrated as follows: in the sub-Hubble scale when $k > \mathcal{H}$, the EM mode functions remain 
in Bunch-Davies vacuum state; and in the super-Hubble scale when $k < \mathcal{H}$, the $A_{\pm}(k,\eta)$ are given by 
Eq.(\ref{superhorizon form 1}). Here $C_i$ are the integration constants which can be determined by matching $A_{\pm}(k,\eta)$ and 
$A_{\pm}'(k,\eta)$ at the transition time of sub-Hubble and super-Hubble regimes, i.e when $k = \mathcal{H}$. If $\eta_{*}$ is the horizon crossing 
instance of the mode $k$, then we have $k\eta_{*} = -(1+\delta)$, where $\delta$ is shown in Eq.(\ref{cosmic Hubble parameter}). 
As a result, the $C_i$ are given by the following expressions \cite{Bamba:2021wyx}, 

\begin{eqnarray}
 C_1&=&\frac{-1}{2\alpha\sqrt{-2k\eta_{*}/\eta_0}}\left[i\pi e^{-ik\eta_{*}}\left\{-\zeta\sqrt{k}\tau_{*}~
 Y_{1+\frac{1}{2\alpha}}\left(-i\frac{\zeta\sqrt{k}}{\alpha}\tau_{*}\right) 
 + k\eta_{*}~Y_{\frac{1}{2\alpha}}\left(-i\frac{\zeta\sqrt{k}}{\alpha}\tau_{*}\right)\right\}\right]~~,\nonumber\\
C_2&=&\frac{-1}{4\alpha\sqrt{-2k\eta_{*}/\eta_0}}\bigg[i\pi e^{-ik\eta_{*}}\bigg\{-\zeta\sqrt{k}\tau_{*}
 \left(J_{-1+\frac{1}{2\alpha}}\left(-i\frac{\zeta\sqrt{k}}{\alpha}\tau_{*}\right) 
 - J_{1+\frac{1}{2\alpha}}\left(-i\frac{\zeta\sqrt{k}}{\alpha}\tau_{*}\right)\right)\nonumber\\
 &+&\left(i - 2k\eta_{*}\right)~J_{\frac{1}{2\alpha}}\left(-i\frac{\zeta\sqrt{k}}{\alpha}\tau_{*}\right)
 \bigg\}\bigg]~~.
 \label{app1-4}
 \end{eqnarray}
 Similarly,
 \begin{eqnarray}
 C_3&=&\frac{-1}{2\alpha\sqrt{-2k\eta_{*}/\eta_0}}\left[-\pi e^{-ik\eta_{*}}\left\{-\zeta\sqrt{k}\tau_{*}~
 Y_{1+\frac{1}{2\alpha}}\left(\frac{\zeta\sqrt{k}}{\alpha}\tau_{*}\right) 
 - ik\eta_{*}~Y_{\frac{1}{2\alpha}}\left(\frac{\zeta\sqrt{k}}{\alpha}\tau_{*}\right)\right\}\right]~~,\nonumber\\
 C_4&=&\frac{-1}{2\alpha\sqrt{-2k\eta_{*}/\eta_0}}\left[\pi e^{-ik\eta_{*}}\left\{-\zeta\sqrt{k}\tau_{*}~
 J_{1+\frac{1}{2\alpha}}\left(\frac{\zeta\sqrt{k}}{\alpha}\tau_{*}\right) 
 - ik\eta_{*}~J_{\frac{1}{2\alpha}}\left(\frac{\zeta\sqrt{k}}{\alpha}\tau_{*}\right)\right\}\right]~.
 \label{app1-5}
\end{eqnarray}
In the above expressions, $\tau_{*} = \left(-\eta_0/\eta_{*}\right)^{\alpha}$ and $\alpha$, $\zeta^2$ are shown earlier in Eq.(\ref{B}). 

\subsection*{Acknowledgments}
TP sincerely acknowledges Sergei D. Odintsov for useful discussions. 
This research was supported in part by the International Centre for Theoretical Sciences (ICTS) 
for the online program - Physics of the Early Universe: ICTS/peu2022/1.


\begin{thebibliography}{99}

%\cite{Bamba:2021wyx}
%\cite{Bamba:2021wyx}
\bibitem{Bamba:2021wyx}
K.~Bamba, S.~D.~Odintsov, T.~Paul and D.~Maity,
%``Helical magnetogenesis with reheating phase from higher curvature coupling and baryogenesis,''
Phys. Dark Univ. \textbf{36} (2022), 101025
doi:10.1016/j.dark.2022.101025
[arXiv:2107.11524 [gr-qc]].
%4 citations counted in INSPIRE as of 08 May 2022

  %\cite{Grasso:2000wj}
\bibitem{Grasso:2000wj}
D.~Grasso and H.~R.~Rubinstein,
%``Magnetic fields in the early universe,''
Phys. Rept. \textbf{348} (2001), 163-266
%doi:10.1016/S0370-1573(00)00110-1
[arXiv:astro-ph/0009061 [astro-ph]].
%792 citations counted in INSPIRE as of 10 Dec 2020

%\cite{Beck:2000dc}
\bibitem{Beck:2000dc}
R.~Beck,
%``Galactic and extragalactic magnetic fields,''
Space Sci. Rev. \textbf{99} (2001), 243-260
%doi:10.1023/A:1013805401252
[arXiv:astro-ph/0012402 [astro-ph]].
%202 citations counted in INSPIRE as of 10 Dec 2020


%\cite{Widrow:2002ud}
\bibitem{Widrow:2002ud}
L.~M.~Widrow,
%``Origin of galactic and extragalactic magnetic fields,''
Rev. Mod. Phys. \textbf{74} (2002), 775-823
%doi:10.1103/RevModPhys.74.775
[arXiv:astro-ph/0207240 [astro-ph]].
%543 citations counted in INSPIRE as of 10 Dec 2020









%\cite{Kulsrud:2007an}
\bibitem{Kulsrud:2007an}
R.~M.~Kulsrud and E.~G.~Zweibel,
%``The Origin of Astrophysical Magnetic Fields,''
Rept. Prog. Phys. \textbf{71} (2008), 0046091
%doi:10.1088/0034-4885/71/4/046901
[arXiv:0707.2783 [astro-ph]].
%112 citations counted in INSPIRE as of 10 Dec 2020

%\cite{Brandenburg:2004jv}
\bibitem{Brandenburg:2004jv}
A.~Brandenburg and K.~Subramanian,
%``Astrophysical magnetic fields and nonlinear dynamo theory,''
Phys. Rept. \textbf{417} (2005), 1-209
%doi:10.1016/j.physrep.2005.06.005
[arXiv:astro-ph/0405052 [astro-ph]].
%539 citations counted in INSPIRE as of 10 Dec 2020


%\cite{Subramanian:2009fu}
\bibitem{Subramanian:2009fu}
K.~Subramanian,
%``Magnetic fields in the early universe,''
Astron. Nachr. \textbf{331} (2010), 110-120
%doi:10.1002/asna.200911312
[arXiv:0911.4771 [astro-ph.CO]].
%95 citations counted in INSPIRE as of 10 Dec 2020



%\cite{Jain:2012ga}
\bibitem{Jain:2012ga}
R.~K.~Jain and M.~S.~Sloth,
%``Consistency relation for cosmic magnetic fields,''
Phys. Rev. D \textbf{86} (2012), 123528
%doi:10.1103/PhysRevD.86.123528
[arXiv:1207.4187 [astro-ph.CO]].
%42 citations counted in INSPIRE as of 10 Dec 2020



%\cite{Durrer:2010mq}
\bibitem{Durrer:2010mq}
R.~Durrer, L.~Hollenstein and R.~K.~Jain,
%``Can slow roll inflation induce relevant helical magnetic fields?,''
JCAP \textbf{03} (2011), 037
%doi:10.1088/1475-7516/2011/03/037
[arXiv:1005.5322 [astro-ph.CO]].
%109 citations counted in INSPIRE as of 10 Dec 2020


%\cite{Kanno:2009ei}
\bibitem{Kanno:2009ei}
S.~Kanno, J.~Soda and M.~a.~Watanabe,
%``Cosmological Magnetic Fields from Inflation and Backreaction,''
JCAP \textbf{12} (2009), 009
%doi:10.1088/1475-7516/2009/12/009
[arXiv:0908.3509 [astro-ph.CO]].
%161 citations counted in INSPIRE as of 10 Dec 2020


%\cite{Campanelli:2008kh}
\bibitem{Campanelli:2008kh}
L.~Campanelli,
%``Helical Magnetic Fields from Inflation,''
Int. J. Mod. Phys. D \textbf{18} (2009), 1395-1411
%doi:10.1142/S0218271809015175
[arXiv:0805.0575 [astro-ph]].
%67 citations counted in INSPIRE as of 10 Dec 2020




%\cite{Demozzi:2009fu}
\bibitem{Demozzi:2009fu}
V.~Demozzi, V.~Mukhanov and H.~Rubinstein,
%``Magnetic fields from inflation?,''
JCAP \textbf{08} (2009), 025
%doi:10.1088/1475-7516/2009/08/025
[arXiv:0907.1030 [astro-ph.CO]].
%234 citations counted in INSPIRE as of 10 Dec 2020

%\cite{Demozzi:2012wh}
\bibitem{Demozzi:2012wh}
V.~Demozzi and C.~Ringeval,
%``Reheating constraints in inflationary magnetogenesis,''
JCAP \textbf{05}, 009 (2012)
%doi:10.1088/1475-7516/2012/05/009
[arXiv:1202.3022 [astro-ph.CO]].
%30 citations counted in INSPIRE as of 21 Apr 2021


  
  
   
   


%\cite{Bamba:2006ga}
\bibitem{Bamba:2006ga}
K.~Bamba and M.~Sasaki,
%``Large-scale magnetic fields in the inflationary universe,''
JCAP \textbf{02} (2007), 030
%doi:10.1088/1475-7516/2007/02/030
[arXiv:astro-ph/0611701 [astro-ph]].
%123 citations counted in INSPIRE as of 10 Dec 2020



%\cite{Kobayashi:2019uqs}
\bibitem{Kobayashi:2019uqs}
T.~Kobayashi and M.~S.~Sloth,
%``Early Cosmological Evolution of Primordial Electromagnetic Fields,''
Phys. Rev. D \textbf{100} (2019) no.2, 023524
%doi:10.1103/PhysRevD.100.023524
[arXiv:1903.02561 [astro-ph.CO]].
%20 citations counted in INSPIRE as of 10 Dec 2020



%\cite{Bamba:2020qdj}
\bibitem{Bamba:2020qdj}
K.~Bamba, E.~Elizalde, S.~D.~Odintsov and T.~Paul,
%``Inflationary magnetogenesis with reheating phase from higher curvature coupling,''
JCAP \textbf{04} (2021), 009
doi:10.1088/1475-7516/2021/04/009
[arXiv:2012.12742 [gr-qc]].
%7 citations counted in INSPIRE as of 17 Jul 2021


%\cite{Maity:2021qps}
\bibitem{Maity:2021qps}
D.~Maity, S.~Pal and T.~Paul,
%``Effective Theory of Inflationary Magnetogenesis and Constraints on Reheating,''
JCAP \textbf{05} (2021), 045
doi:10.1088/1475-7516/2021/05/045
[arXiv:2103.02411 [hep-th]].
%2 citations counted in INSPIRE as of 17 Jul 2021



%\cite{Haque:2020bip}
\bibitem{Haque:2020bip}
M.~R.~Haque, D.~Maity and S.~Pal,
%``Probing the reheating phase through primordial magnetic field and CMB,''
[arXiv:2012.10859 [hep-th]].
%1 citations counted in INSPIRE as of 28 Jan 2021











%\cite{Ratra:1991bn}
\bibitem{Ratra:1991bn}
B.~Ratra,
%``Cosmological 'seed' magnetic field from inflation,''
Astrophys. J. Lett. \textbf{391} (1992), L1-L4
%oi:10.1086/186384
%652 citations counted in INSPIRE as of 10 Dec 2020

%\cite{Ade:2015cva}
\bibitem{Ade:2015cva}
P.~A.~R.~Ade \textit{et al.} [Planck],
%``Planck 2015 results. XIX. Constraints on primordial magnetic fields,''
Astron. Astrophys. \textbf{594} (2016), A19
%doi:10.1051/0004-6361/201525821
[arXiv:1502.01594 [astro-ph.CO]].
%248 citations counted in INSPIRE as of 10 Dec 2020


%\cite{Chowdhury:2018mhj}
\bibitem{Chowdhury:2018mhj}
D.~Chowdhury, L.~Sriramkumar and M.~Kamionkowski,
%``Enhancing the cross-correlations between magnetic fields and scalar perturbations through parity violation,''
JCAP \textbf{10} (2018), 031
%doi:10.1088/1475-7516/2018/10/031
[arXiv:1807.07477 [astro-ph.CO]].
%6 citations counted in INSPIRE as of 10 Dec 2020





%\cite{Turner:1987bw}
\bibitem{Turner:1987bw}
M.~S.~Turner and L.~M.~Widrow,
%``Inflation Produced, Large Scale Magnetic Fields,''
Phys. Rev. D \textbf{37} (1988), 2743
%doi:10.1103/PhysRevD.37.2743
%781 citations counted in INSPIRE as of 10 Dec 2020


%\cite{Tripathy:2021sfb}
\bibitem{Tripathy:2021sfb}
S.~Tripathy, D.~Chowdhury, R.~K.~Jain and L.~Sriramkumar,
%``Challenges in the choice of the nonconformal coupling function in inflationary magnetogenesis,''
Phys. Rev. D \textbf{105} (2022) no.6, 063519
doi:10.1103/PhysRevD.105.063519
[arXiv:2111.01478 [astro-ph.CO]].
%1 citations counted in INSPIRE as of 06 Apr 2022




%\cite{Ferreira:2013sqa}
\bibitem{Ferreira:2013sqa}
R.~J.~Z.~Ferreira, R.~K.~Jain and M.~S.~Sloth,
%``Inflationary magnetogenesis  without the strong coupling problem,''
JCAP \textbf{10} (2013), 004
%doi:10.1088/1475-7516/2013/10/004
[arXiv:1305.7151 [astro-ph.CO]].
%94 citations counted in INSPIRE as of 10 Dec 2020

%\cite{Atmjeet:2014cxa}
\bibitem{Atmjeet:2014cxa}
K.~Atmjeet, T.~R.~Seshadri and K.~Subramanian,
%``Helical cosmological magnetic fields from extra-dimensions,''
Phys. Rev. D \textbf{91} (2015), 103006
%doi:10.1103/PhysRevD.91.103006
[arXiv:1409.6840 [astro-ph.CO]].
%12 citations counted in INSPIRE as of 10 Dec 2020


%\cite{Kushwaha:2020nfa}
\bibitem{Kushwaha:2020nfa}
A.~Kushwaha and S.~Shankaranarayanan,
%``Helical magnetic fields from Riemann coupling,''
Phys. Rev. D \textbf{102} (2020) no.10, 103528
doi:10.1103/PhysRevD.102.103528
[arXiv:2008.10825 [gr-qc]].
%4 citations counted in INSPIRE as of 18 Sep 2021


%\cite{Gasperini:1995dh}
\bibitem{Gasperini:1995dh}
M.~Gasperini, M.~Giovannini and G.~Veneziano,
%``Primordial magnetic fields from string cosmology,''
Phys. Rev. Lett. \textbf{75} (1995), 3796-3799
doi:10.1103/PhysRevLett.75.3796
[arXiv:hep-th/9504083 [hep-th]].
%310 citations counted in INSPIRE as of 21 May 2022


%\cite{Giovannini:2021thf}
\bibitem{Giovannini:2021thf}
M.~Giovannini,
%``Large-scale gauge spectra and pseudoscalar couplings,''
Phys. Rev. D \textbf{104} (2021) no.12, 123509
doi:10.1103/PhysRevD.104.123509
[arXiv:2106.14927 [hep-th]].
%4 citations counted in INSPIRE as of 21 May 2022


%\cite{Giovannini:2021xbi}
\bibitem{Giovannini:2021xbi}
M.~Giovannini,
%``Baryogenesis, magnetogenesis and the strength of anomalous interactions,''
Eur. Phys. J. C \textbf{81} (2021) no.6, 503
doi:10.1140/epjc/s10052-021-09282-7
[arXiv:2103.04137 [astro-ph.CO]].
%5 citations counted in INSPIRE as of 21 May 2022





\bibitem{Adshead:2015pva}
P.~Adshead, J.~T.~Giblin, T.~R.~Scully and E.~I.~Sfakianakis,
%``Gauge-preheating and the end of axion inflation,''
JCAP \textbf{12} (2015), 034
%doi:10.1088/1475-7516/2015/12/034
[arXiv:1502.06506 [astro-ph.CO]].












%\cite{Caprini:2014mja}
\bibitem{Caprini:2014mja}
C.~Caprini and L.~Sorbo,
%``Adding helicity to inflationary magnetogenesis,''
JCAP \textbf{10} (2014), 056
%doi:10.1088/1475-7516/2014/10/056
[arXiv:1407.2809 [astro-ph.CO]].
%108 citations counted in INSPIRE as of 10 Dec 2020


%\cite{Kobayashi:2014sga}
\bibitem{Kobayashi:2014sga}
T.~Kobayashi,
%``Primordial Magnetic Fields from the Post-Inflationary Universe,''
JCAP \textbf{05} (2014), 040
%doi:10.1088/1475-7516/2014/05/040
[arXiv:1403.5168 [astro-ph.CO]].
%45 citations counted in INSPIRE as of 10 Dec 2020


%\cite{Atmjeet:2013yta}
\bibitem{Atmjeet:2013yta}
K.~Atmjeet, I.~Pahwa, T.~R.~Seshadri and K.~Subramanian,
%``Cosmological Magnetogenesis From Extra-dimensional Gauss Bonnet Gravity,''
Phys. Rev. D \textbf{89} (2014) no.6, 063002
%doi:10.1103/PhysRevD.89.063002
[arXiv:1312.5815 [astro-ph.CO]].
%18 citations counted in INSPIRE as of 10 Dec 2020


%\cite{Fujita:2015iga}
\bibitem{Fujita:2015iga}
T.~Fujita, R.~Namba, Y.~Tada, N.~Takeda and H.~Tashiro,
%``Consistent generation of magnetic fields in axion inflation models,''
JCAP \textbf{05} (2015), 054
%doi:10.1088/1475-7516/2015/05/054
[arXiv:1503.05802 [astro-ph.CO]].
%70 citations counted in INSPIRE as of 10 Dec 2020


%\cite{Campanelli:2015jfa}
\bibitem{Campanelli:2015jfa}
L.~Campanelli,
%``Lorentz-violating inflationary magnetogenesis,''
Eur. Phys. J. C \textbf{75} (2015) no.6, 278
%doi:10.1140/epjc/s10052-015-3510-x
[arXiv:1503.07415 [gr-qc]].
%23 citations counted in INSPIRE as of 10 Dec 2020


%\cite{Tasinato:2014fia}
\bibitem{Tasinato:2014fia}
G.~Tasinato,
%``A scenario for inflationary magnetogenesis without strong coupling problem,''
JCAP \textbf{03} (2015), 040
%doi:10.1088/1475-7516/2015/03/040
[arXiv:1411.2803 [hep-th]].
%31 citations counted in INSPIRE as of 10 Dec 2020


%\cite{Nandi:2021lpf}
\bibitem{Nandi:2021lpf}
D.~Nandi,
%``Inflationary magnetogenesis: solving the strong coupling and its non-Gaussian signatures,''
JCAP \textbf{08} (2021), 039
doi:10.1088/1475-7516/2021/08/039
[arXiv:2103.03159 [astro-ph.CO]].
%2 citations counted in INSPIRE as of 23 Apr 2022





%\cite{Frion:2020bxc}
\bibitem{Frion:2020bxc}
E.~Frion, N.~Pinto-Neto, S.~D.~P.~Vitenti and S.~E.~Perez Bergliaffa,
%``Primordial Magnetogenesis in a Bouncing Universe,''
Phys. Rev. D \textbf{101} (2020) no.10, 103503
%doi:10.1103/PhysRevD.101.103503
[arXiv:2004.07269 [gr-qc]].
%3 citations counted in INSPIRE as of 10 Dec 2020


%\cite{Koley:2016jdw}
\bibitem{Koley:2016jdw}
R.~Koley and S.~Samtani,
%``Magnetogenesis in Matter - Ekpyrotic Bouncing Cosmology,''
JCAP \textbf{04} (2017), 030
%doi:10.1088/1475-7516/2017/04/030
[arXiv:1612.08556 [gr-qc]].
%10 citations counted in INSPIRE as of 10 Dec 2020



%\cite{Qian:2016lbf}
\bibitem{Qian:2016lbf}
P.~Qian, Y.~F.~Cai, D.~A.~Easson and Z.~K.~Guo,
%``Magnetogenesis in bouncing cosmology,''
Phys. Rev. D \textbf{94} (2016) no.8, 083524
%doi:10.1103/PhysRevD.94.083524
[arXiv:1607.06578 [gr-qc]].
%13 citations counted in INSPIRE as of 10 Dec 2020





 \bibitem{guth}
A.H. Guth;  Phys.Rev. D23 347-356 (1981).

  %\cite{Linde:2005ht}
\bibitem{Linde:2005ht}
  A.~D.~Linde,
  %``Particle physics and inflationary cosmology,''
  Contemp.\ Concepts Phys.\  {\bf 5} (1990) 1
  [hep-th/0503203].
  %%CITATION = HEP-TH/0503203;%%
  %676 citations counted in INSPIRE as of 21 Oct 2019


 %\cite{Langlois:2004de}
\bibitem{Langlois:2004de}
  D.~Langlois,
  %``Inflation, quantum fluctuations and cosmological perturbations,''
  hep-th/0405053.
  %%CITATION = HEP-TH/0405053;%%
  %61 citations counted in INSPIRE as of 21 Oct 2019


  %\cite{Riotto:2002yw}
\bibitem{Riotto:2002yw}
  A.~Riotto,
  %``Inflation and the theory of cosmological perturbations,''
  ICTP Lect.\ Notes Ser.\  {\bf 14} (2003) 317
  [hep-ph/0210162].
  %%CITATION = HEP-PH/0210162;%%
  %313 citations counted in INSPIRE as of 21 Oct 2019




%\cite{Baumann:2009ds}
\bibitem{Baumann:2009ds}
D.~Baumann,
%``Inflation,''
%doi:10.1142/9789814327183 0010
[arXiv:0907.5424 [hep-th]].
%729 citations counted in INSPIRE as of 24 Aug 2020




%\cite{Dai:2014jja}
\bibitem{Dai:2014jja}
L.~Dai, M.~Kamionkowski and J.~Wang,
%``Reheating constraints to inflationary models,''
Phys. Rev. Lett. \textbf{113} (2014), 041302
%doi:10.1103/PhysRevLett.113.041302
[arXiv:1404.6704 [astro-ph.CO]].
%149 citations counted in INSPIRE as of 10 Dec 2020

%\cite{Cook:2015vqa}
\bibitem{Cook:2015vqa}
J.~L.~Cook, E.~Dimastrogiovanni, D.~A.~Easson and L.~M.~Krauss,
%``Reheating predictions in single field inflation,''
JCAP \textbf{04} (2015), 047
%doi:10.1088/1475-7516/2015/04/047
[arXiv:1502.04673 [astro-ph.CO]].
%108 citations counted in INSPIRE as of 10 Dec 2020


%\cite{Albrecht:1982mp}
\bibitem{Albrecht:1982mp}
A.~Albrecht, P.~J.~Steinhardt, M.~S.~Turner and F.~Wilczek,
%``Reheating an Inflationary Universe,''
Phys. Rev. Lett. \textbf{48} (1982), 1437
%doi:10.1103/PhysRevLett.48.1437
%497 citations counted in INSPIRE as of 10 Dec 2020


%\cite{Ellis:2015pla}
\bibitem{Ellis:2015pla}
J.~Ellis, M.~A.~G.~Garcia, D.~V.~Nanopoulos and K.~A.~Olive,
%``Calculations of Inflaton Decays and Reheating: with Applications to No-Scale Inflation Models,''
JCAP \textbf{07} (2015), 050
%doi:10.1088/1475-7516/2015/07/050
[arXiv:1505.06986 [hep-ph]].
%58 citations counted in INSPIRE as of 10 Dec 2020


%\cite{Ueno:2016dim}
\bibitem{Ueno:2016dim}
Y.~Ueno and K.~Yamamoto,
%``Constraints on $\alpha$-attractor inflation and reheating,''
Phys. Rev. D \textbf{93} (2016) no.8, 083524
%doi:10.1103/PhysRevD.93.083524
[arXiv:1602.07427 [astro-ph.CO]].
%47 citations counted in INSPIRE as of 10 Dec 2020


%\cite{Eshaghi:2016kne}
\bibitem{Eshaghi:2016kne}
M.~Eshaghi, M.~Zarei, N.~Riazi and A.~Kiasatpour,
%``CMB and reheating constraints to $\alpha$-attractor inflationary models,''
Phys. Rev. D \textbf{93} (2016) no.12, 123517
%doi:10.1103/PhysRevD.93.123517
[arXiv:1602.07914 [astro-ph.CO]].
%28 citations counted in INSPIRE as of 10 Dec 2020


%\cite{Maity:2018qhi}
\bibitem{Maity:2018qhi}
D.~Maity and P.~Saha,
%``(P)reheating after minimal Plateau Inflation and constraints from CMB,''
JCAP \textbf{07} (2019), 018
%doi:10.1088/1475-7516/2019/07/018
[arXiv:1811.11173 [astro-ph.CO]].
%8 citations counted in INSPIRE as of 10 Dec 2020

%\cite{Haque:2021dha}
\bibitem{Haque:2021dha}
M.~R.~Haque, D.~Maity, T.~Paul and L.~Sriramkumar,
%``Decoding the phases of early and late time reheating through imprints on primordial gravitational waves,''
Phys. Rev. D \textbf{104} (2021) no.6, 063513
doi:10.1103/PhysRevD.104.063513
[arXiv:2105.09242 [astro-ph.CO]].
%4 citations counted in INSPIRE as of 12 Mar 2022


%\cite{DiMarco:2017zek}
\bibitem{DiMarco:2017zek}
A.~Di Marco, P.~Cabella and N.~Vittorio,
%``Constraining the general reheating phase in the $\alpha$-attractor inflationary cosmology,''
Phys. Rev. D \textbf{95} (2017) no.10, 103502
%doi:10.1103/PhysRevD.95.103502
[arXiv:1705.04622 [astro-ph.CO]].
%9 citations counted in INSPIRE as of 10 Dec 2020


%\cite{Drewes:2017fmn}
\bibitem{Drewes:2017fmn}
M.~Drewes, J.~U.~Kang and U.~R.~Mun,
%``CMB constraints on the inflaton couplings and reheating temperature in $\alpha$-attractor inflation,''
JHEP \textbf{11} (2017), 072
%doi:10.1007/JHEP11(2017)072
[arXiv:1708.01197 [astro-ph.CO]].
%15 citations counted in INSPIRE as of 10 Dec 2020


%\cite{Nojiri:2005vv}
\bibitem{Nojiri:2005vv}
S.~Nojiri, S.~D.~Odintsov and M.~Sasaki,
%``Gauss-Bonnet dark energy,''
Phys. Rev. D \textbf{71} (2005), 123509
doi:10.1103/PhysRevD.71.123509
[arXiv:hep-th/0504052 [hep-th]].
%571 citations counted in INSPIRE as of 18 Jul 2021

%\cite{Li:2007jm}
\bibitem{Li:2007jm}
B.~Li, J.~D.~Barrow and D.~F.~Mota,
%``The Cosmology of Modified Gauss-Bonnet Gravity,''
Phys. Rev. D \textbf{76} (2007), 044027
doi:10.1103/PhysRevD.76.044027
[arXiv:0705.3795 [gr-qc]].
%197 citations counted in INSPIRE as of 02 May 2020

%\cite{Carter:2005fu}
\bibitem{Carter:2005fu}
B.~M.~Carter and I.~P.~Neupane,
%``Towards inflation and dark energy cosmologies from modified Gauss-Bonnet theory,''
JCAP \textbf{06} (2006), 004
doi:10.1088/1475-7516/2006/06/004
[arXiv:hep-th/0512262 [hep-th]].
%131 citations counted in INSPIRE as of 02 May 2020

%\cite{Nojiri:2019dwl}
\bibitem{Nojiri:2019dwl}
S.~Nojiri, S.~Odintsov, V.~Oikonomou, N.~Chatzarakis and T.~Paul,
%``Viable inflationary models in a ghost-free Gauss–Bonnet theory of gravity,''
Eur. Phys. J. C \textbf{79} (2019) no.7, 565
doi:10.1140/epjc/s10052-019-7080-1
[arXiv:1907.00403 [gr-qc]].
%6 citations counted in INSPIRE as of 12 Jun 2020



%\cite{Cognola:2006eg}
\bibitem{Cognola:2006eg}
G.~Cognola, E.~Elizalde, S.~Nojiri, S.~D.~Odintsov and S.~Zerbini,
%``Dark energy in modified Gauss-Bonnet gravity: Late-time acceleration and the hierarchy problem,''
Phys. Rev. D \textbf{73} (2006), 084007
doi:10.1103/PhysRevD.73.084007
[arXiv:hep-th/0601008 [hep-th]].
%506 citations counted in INSPIRE as of 02 May 2020

%\cite{Chakraborty:2018scm}
\bibitem{Chakraborty:2018scm}
S.~Chakraborty, T.~Paul and S.~SenGupta,
%``Inflation driven by Einstein-Gauss-Bonnet gravity,''
Phys. Rev. D \textbf{98} (2018) no.8, 083539
doi:10.1103/PhysRevD.98.083539
[arXiv:1804.03004 [gr-qc]].
%56 citations counted in INSPIRE as of 06 Apr 2022

%\cite{Elizalde:2020zcb}
\bibitem{Elizalde:2020zcb}
E.~Elizalde, S.~D.~Odintsov, V.~K.~Oikonomou and T.~Paul,
%``Extended matter bounce scenario in ghost free $f(R,\mathcal{G})$ gravity compatible with GW170817,''
Nucl. Phys. B \textbf{954} (2020), 114984
doi:10.1016/j.nuclphysb.2020.114984
[arXiv:2003.04264 [gr-qc]].
%32 citations counted in INSPIRE as of 06 Apr 2022

%\cite{Nojiri:2022xdo}
\bibitem{Nojiri:2022xdo}
S.~Nojiri, S.~D.~Odintsov and T.~Paul,
%``Towards a smooth unification from an ekpyrotic bounce to the dark energy era,''
Phys. Dark Univ. \textbf{35} (2022), 100984
doi:10.1016/j.dark.2022.100984
[arXiv:2202.02695 [gr-qc]].
%1 citations counted in INSPIRE as of 23 Apr 2022


%\cite{Odintsov:2022unp}
\bibitem{Odintsov:2022unp}
S.~D.~Odintsov and T.~Paul,
%``Bounce universe with finite-time singularity,''
[arXiv:2205.09447 [gr-qc]].
%0 citations counted in INSPIRE as of 21 May 2022


%\cite{Planck:2018jri}
\bibitem{Planck:2018jri}
Y.~Akrami \textit{et al.} [Planck],
%``Planck 2018 results. X. Constraints on inflation,''
Astron. Astrophys. \textbf{641} (2020), A10
doi:10.1051/0004-6361/201833887
[arXiv:1807.06211 [astro-ph.CO]].
%1403 citations counted in INSPIRE as of 17 Jul 2021

%\cite{Burgess:2009ea}
\bibitem{Burgess:2009ea}
C.~P.~Burgess, H.~M.~Lee and M.~Trott,
%``Power-counting and the Validity of the Classical Approximation During Inflation,''
JHEP \textbf{09} (2009), 103
doi:10.1088/1126-6708/2009/09/103
[arXiv:0902.4465 [hep-ph]].
%327 citations counted in INSPIRE as of 18 Sep 2021


%\cite{Hertzberg:2010dc}
\bibitem{Hertzberg:2010dc}
M.~P.~Hertzberg,
%``On Inflation with Non-minimal Coupling,''
JHEP \textbf{11} (2010), 023
doi:10.1007/JHEP11(2010)023
[arXiv:1002.2995 [hep-ph]].
%248 citations counted in INSPIRE as of 18 Sep 2021


%\cite{Bezrukov:2010jz}
\bibitem{Bezrukov:2010jz}
F.~Bezrukov, A.~Magnin, M.~Shaposhnikov and S.~Sibiryakov,
%``Higgs inflation: consistency and generalisations,''
JHEP \textbf{01} (2011), 016
doi:10.1007/JHEP01(2011)016
[arXiv:1008.5157 [hep-ph]].
%448 citations counted in INSPIRE as of 18 Sep 2021


%\cite{Ferreira:2015omg}
\bibitem{Ferreira:2015omg}
R.~Z.~Ferreira, J.~Ganc, J.~Nore\~na and M.~S.~Sloth,
%``On the validity of the perturbative description of axions during inflation,''
JCAP \textbf{04} (2016), 039
[erratum: JCAP \textbf{10} (2016), E01]
doi:10.1088/1475-7516/2016/04/039
[arXiv:1512.06116 [astro-ph.CO]].
%54 citations counted in INSPIRE as of 18 Sep 2021

%\cite{Fujita:2013qxa}
\bibitem{Fujita:2013qxa}
T.~Fujita and S.~Yokoyama,
%``Higher order statistics of curvature perturbations in IFF model and its Planck constraints,''
JCAP \textbf{09} (2013), 009
doi:10.1088/1475-7516/2013/09/009
[arXiv:1306.2992 [astro-ph.CO]].
%60 citations counted in INSPIRE as of 08 Feb 2021




%\cite{Barnaby:2012tk}
\bibitem{Barnaby:2012tk}
N.~Barnaby, R.~Namba and M.~Peloso,
%``Observable non-gaussianity from gauge field production in slow roll inflation, and a challenging connection with magnetogenesis,''
Phys. Rev. D \textbf{85} (2012), 123523
doi:10.1103/PhysRevD.85.123523
[arXiv:1202.1469 [astro-ph.CO]].
%142 citations counted in INSPIRE as of 08 Feb 2021




%\cite{Bamba:2014vda}
\bibitem{Bamba:2014vda}
K.~Bamba,
%``Generation of large-scale magnetic fields, non-Gaussianity, and primordial gravitational waves in inflationary cosmology,''
Phys. Rev. D \textbf{91} (2015), 043509
doi:10.1103/PhysRevD.91.043509
[arXiv:1411.4335 [astro-ph.CO]].
%20 citations counted in INSPIRE as of 08 Feb 2021




%\cite{Suyama:2012wh}
\bibitem{Suyama:2012wh}
T.~Suyama and J.~Yokoyama,
%``Metric perturbation from inflationary magnetic field and generic bound on inflation models,''
Phys. Rev. D \textbf{86} (2012), 023512
doi:10.1103/PhysRevD.86.023512
[arXiv:1204.3976 [astro-ph.CO]].
%38 citations counted in INSPIRE as of 08 Feb 2021


%\cite{Shankaranarayanan:2004iq}
\bibitem{Shankaranarayanan:2004iq}
S.~Shankaranarayanan and L.~Sriramkumar,
%``Trans-Planckian corrections to the primordial spectrum in the infrared and the ultraviolet,''
Phys. Rev. D \textbf{70} (2004), 123520
doi:10.1103/PhysRevD.70.123520
[arXiv:hep-th/0403236 [hep-th]].
%28 citations counted in INSPIRE as of 12 May 2022


%\cite{delCampo:2015wma}
\bibitem{delCampo:2015wma}
S.~del Campo, C.~Gonzalez and R.~Herrera,
%``Power law inflation with a non-minimally coupled scalar field in light of Planck 2015 data: the exact versus slow roll results,''
Astrophys. Space Sci. \textbf{358} (2015) no.2, 31
doi:10.1007/s10509-015-2414-4
[arXiv:1501.05697 [gr-qc]].
%7 citations counted in INSPIRE as of 12 May 2022


%\cite{Sharma:2022tce}
\bibitem{Sharma:2022tce}
A.~K.~Sharma and M.~M.~Verma,
%``Power-law Inflation in the $f$(R) Gravity,''
Astrophys. J. \textbf{926} (2022) no.1, 29
doi:10.3847/1538-4357/ac3ed7
%3 citations counted in INSPIRE as of 12 May 2022


%\cite{Pham:2021fjj}
\bibitem{Pham:2021fjj}
T.~M.~Pham, D.~H.~Nguyen and T.~Q.~Do,
%``k-Gauss-Bonnet inflation,''
[arXiv:2107.05926 [gr-qc]].
%0 citations counted in INSPIRE as of 12 May 2022


%\cite{Guendelman:1991se}
\bibitem{Guendelman:1991se}
E.~I.~Guendelman and D.~A.~Owen,
%``Axion driven baryogenesis,''
Phys. Lett. B \textbf{276} (1992), 108-114
doi:10.1016/0370-2693(92)90548-I
%14 citations counted in INSPIRE as of 08 May 2022



%\cite{Anber:2006xt}
\bibitem{Anber:2006xt}
M.~M.~Anber and L.~Sorbo,
%``N-flationary magnetic fields,''
JCAP \textbf{10} (2006), 018
doi:10.1088/1475-7516/2006/10/018
[arXiv:astro-ph/0606534 [astro-ph]].
%186 citations counted in INSPIRE as of 07 May 2022


%\cite{Barnaby:2011vw}
\bibitem{Barnaby:2011vw}
N.~Barnaby, R.~Namba and M.~Peloso,
%``Phenomenology of a Pseudo-Scalar Inflaton: Naturally Large Nongaussianity,''
JCAP \textbf{04} (2011), 009
doi:10.1088/1475-7516/2011/04/009
[arXiv:1102.4333 [astro-ph.CO]].
%192 citations counted in INSPIRE as of 07 May 2022


%\cite{Peloso:2016gqs}
\bibitem{Peloso:2016gqs}
M.~Peloso, L.~Sorbo and C.~Unal,
%``Rolling axions during inflation: perturbativity and signatures,''
JCAP \textbf{09} (2016), 001
doi:10.1088/1475-7516/2016/09/001
[arXiv:1606.00459 [astro-ph.CO]].
%83 citations counted in INSPIRE as of 07 May 2022


%\cite{Adshead:2016iae}
\bibitem{Adshead:2016iae}
P.~Adshead, J.~T.~Giblin, T.~R.~Scully and E.~I.~Sfakianakis,
%``Magnetogenesis from axion inflation,''
JCAP \textbf{10} (2016), 039
doi:10.1088/1475-7516/2016/10/039
[arXiv:1606.08474 [astro-ph.CO]].
%110 citations counted in INSPIRE as of 07 May 2022



%\cite{Fujita:2019pmi}
\bibitem{Fujita:2019pmi}
T.~Fujita and R.~Durrer,
%``Scale-invariant Helical Magnetic Fields from Inflation,''
JCAP \textbf{09} (2019), 008
doi:10.1088/1475-7516/2019/09/008
[arXiv:1904.11428 [astro-ph.CO]].
%31 citations counted in INSPIRE as of 07 May 2022

\end{thebibliography}
\end{document}